\documentclass[sigconf]{acmart}

\AtBeginDocument{%
  \providecommand\BibTeX{{%
    \normalfont B\kern-0.5em{\scshape i\kern-0.25em b}\kern-0.8em\TeX}}}

\setcopyright{acmcopyright}
\copyrightyear{2018}
\acmYear{2018}
\acmDOI{10.1145/1122445.1122456}

\usepackage{booktabs} 
\usepackage{url}

\usepackage{caption}
\usepackage[all]{nowidow}
\usepackage{array}
\usepackage{arydshln}
\usepackage{tabularx}
\usepackage{multirow}
\usepackage{arydshln}
\newcolumntype{L}[1]{>{\raggedright\let\newline\\\arraybackslash\hspace{0pt}}m{#1}}
\newcolumntype{C}[1]{>{\centering\let\newline\\\arraybackslash\hspace{0pt}}m{#1}}
\newcolumntype{R}[1]{>{\raggedleft\let\newline\\\arraybackslash\hspace{0pt}}m{#1}}

\def\authnotes{1}
\newcounter{notectr}[section]

\newcommand{\newedits}[1]{\textcolor{black}{#1}}

\newcommand{\jrz}[1]{\textcolor{black}{#1}}

\newcommand{\cnote}[1]{\ifnum\authnotes=1 \textcolor{blue}{\note{Comment:}{#1}}\fi}

\copyrightyear{2026}
\acmYear{2026}
\setcopyright{cc}
\setcctype{by}
\acmConference[CHI '26]{Proceedings of the 2026 CHI Conference on Human Factors in Computing Systems}{April 13--17, 2026}{Barcelona, Spain}
\acmBooktitle{Proceedings of the 2026 CHI Conference on Human Factors in Computing Systems (CHI '26), April 13--17, 2026, Barcelona, Spain}
\acmPrice{}
\acmDOI{10.1145/3772318.3790387}
\acmISBN{979-8-4007-2278-3/2026/04}

\begin{document}



\title[Narrative in Artistic Performances]{Embodying Facts, Figures, and Faiths in Narrative Artistic Performances in Rural Bangladesh}



\author{Sharifa Sultana}
\affiliation{
  \institution{University of Illinois Urbana-Champaign}
  \city{Urbana}
  \state{Illinois}
  \country{USA}}
\email{sharifas@illinois.edu}

\author{Zinnat Sultana}
\affiliation{%
  \institution{S.M.R. Law College}
  \country{Jessore, Bangladesh}}
\email{zinnat1409@gmail.com}

\author{Jeffrey M. Rzeszotarski}
\affiliation{
  \institution{Loyola University Maryland}
  \city{Baltimore}
  \state{Maryland}
  \country{USA}}
\email{jeff.rzeszotarski@gmail.com}

\author{Syed Ishtiaque Ahmed}
\affiliation{
  \institution{University of Toronto}
  \city{Toronto}
  \state{Ontario}
  \country{Canada}}
\email{ishtiaque@cs.toronto.edu}

\renewcommand{\shortauthors}{Sultana et al.}

\begin{abstract}
There is an increasing interest in telling serious stories with data. Designers organize information, construct narratives, and present findings to inform audiences. However, many of these practices emerge from modern information visualization rhetoric and ethical frameworks which may marginalize communities with low digital and media literacy. In a ten-month-long ethnographic study in three Bangladeshi villages, we investigated how these communities use entertainment and cultural practices, namely \textit{Puthi}, \textit{Bhandari Gaan}, and \textit{Pot music}, to instruct, communicate traditional moral lessons and recall history. We found that these communities embrace polyvocality and multiple ethical frameworks in their performances, construct narratives combining factuality, emotionality, and aesthetics, and adapt their performances to changing technology and audience needs. Our findings provide HCI, visualization, and ethical data practitioners with implications for the design of accessible and culturally appropriate ways of presenting data narratives in data-driven systems.


\end{abstract}







\settopmatter{printfolios=true}

\maketitle

\section{Introduction}
Information gathering, curation, and storytelling are active and ongoing practices across domains such as journalism, science, and education. 
They involve identifying needed information and acquiring it from various sources; removing errors and inconsistencies and transforming it into an acceptable format; creating systems for storing and retrieving; providing users with the tools and information they need to find and use it; and preserving and maintaining it so that it is kept safe and accessible over time \cite{cragin2007educational, gold2010data}. High-quality information stewardship requires not only great care and accountability, but also involvement of both the local generators and consumers of the information. 
Historically, information stewardship processes have been led by government agencies, research institutions, national and international consensus, and large companies, \newedits{as well as local leaders (e.g., chiefs, kings, community leaders and other local structures), depending on the part of the world that one is from} \cite{DataCura83:online, CENDIDat6:online, antognoli2020reproducibility, BTSDataS30:online, Aboutthe22:online, DataCata85:online, GooglePu52:online, PagePubl73:online}. Such entities in power often risked privileging biased perspectives and manipulating records\footnote{\jrz{There are also many cases of information infrastructures explicitly employed for malign purposes by power, such as early computerized record-keeping in Nazi Germany \cite{luebke1994locating}. In this paper, we specifically focus on the ways that mainstream practices may neglect to consider local perspectives, rather than the ways they can be explicitly used to harm.}}, discarding the voices of marginalized ones \cite{folbre1989women, wachter2000fifth, grundmeier2015imputing}. Examples can be found in historical records assembled by colonial governments \cite{burton1994burdens, dirks2006scandal}, reports on struggles of the working class recorded by elites \cite{prentice2018health, yuan2022occupational}, and racial history recorded by dominant groups \cite{williams2015racial, donovan2019toward}. 

In many cases such marginalized groups had limited access to mainstream communication channels, and therefore developed their own ways to communicate with fellow community members in contextual and situated manners, which are rarely part of mainstream history recorded by accountable entities \cite{prasad2006unravelling, rycroft2006santalism, toffin2009janajati}. Due to this disconnect between the origins and continued reification of official recordkeeping as part of imbalanced power relationships, there is a high risk that new efforts using this infrastructure may unwittingly continue to encode biases and further marginalize communities. \jrz{This has led to growing interest in academe and policy-making related to issues such as data sovereignty \cite{walter2021indigenous}.} With technology now irreversibly linked to information-keeping, this disconnect is a major yet understudied concern at the intersection of human-computer interaction (HCI), data science, and critical data studies. 

Our research engages with this agenda through a ten-month-long ethnographic study with rural Jashorian people in Bangladesh, a developing country in South Asia. This area had been under Middle Eastern, British, and Pakistani \jrz{colonial dominion} for several hundred years until 1971. Thus, a significant portion of Bangladeshi history was documented by external sources. Locals’ voices are stored and maintained through traditional performances of myths, folktales, and historical events, among other alternative and informal information practices (e.g., crafting artifacts). In our fieldwork, we engaged with people \newedits{(n=76 in interviews and focus groups and more than 100 in observation sessions)} in three villages in Jashore and investigated their cultural performances associated with local history and moral lessons. These individuals not only perform, but also play a role in educating the villagers on how they interpret and receive information beyond the scope of individual teachings. 
We solicited answers to the following research questions: 

\begin{enumerate}
    \item[\textbf{RQ1}] \jrz{What sustained, major cultural performances and traditional practices involve local history and morals in situated storytelling? }
    \item[\textbf{RQ2}] \jrz{How are the historical accounts, myths, and folktales in these practices passed down over generations?} 
    \item[\textbf{RQ3}] \jrz{How are narratives constructed, and what human factors and material practices are involved in performances?}
    \item[\textbf{RQ4}] How do artists adapt performances to the \jrz{changing landscape of information and communication technologies?} 
\end{enumerate}

We found that communities sustain their histories in cultural performances like \textit{Puthi}, \textit{Bhandari Gaan}, and \textit{Pot music}, which integrate signals and embodied practices embedded in local contexts and trace through lineages of marginalized communities’ collected knowledge. Our deeper investigation revealed that narratives encoded in the performances involve both direct conveyance of information as well as emotion and aesthetics so that the communication is culturally embedded and situated. 
Additionally, we found that the performers adapt their performances to significant changes in society, (inter)national disasters or crises, and political events. Moreover, we observe how material practices in performances and visuals used in constructing narratives changed with the advancement of technology, \jrz{leading to a rich interplay between tradition and technology} in methods of rural storytelling. 

Our paper makes four key contributions to HCI, critical data studies, and information science. First, we provide a detailed understanding of three rural Bangladeshi performances --- Puthi, Bhandari Gaan, and Pot music --- that these communities use as culturally situated forms of \textbf{storytelling and information sharing} in community-driven stewardship. Second, we explain how rural communities rely on \textbf{visual conventions, embodied gestures, and material practices} to enact narrative forms that preserve voices and perspectives missing from conventional historical records. These are enacted in conversation with significant events and changing technology. Third, we show that these storytelling practices operate through an \textbf{intersection of factuality, emotionality, and aesthetics}, constructing narratives that are deeply contextual, situated, and polyvocal. Fourth, building on our findings and literature, we suggest that designers engage with such \textbf{alternative storytelling practices} to develop more inclusive, accessible, and culturally appropriate ways of presenting data narratives.

\section{Related Work}

\subsection{Record-keeping, Preservation, and HCI}
Historically, managing information has meant collecting, organizing, and preserving records for present and future use in scholarship, science, and education \cite{shreeves2008introduction}. We extend this view to record-keeping, where information includes news, accounts, instructions, teachings, and stories transmitted within communities. Such practices ensure accessibility, transparency, and reuse \cite{johnston2018important, weber2012current, palmer2013foundations}. Historically, record-keepers maintained archives of documents, libraries of manuscripts, collections of specimens, and census records \cite{SignsonT9:online, huylebrouck2019missing}. Their roles went beyond administration: in Ancient Egypt, scribes preserved ritual knowledge and gained status through literacy \cite{baines1983literacy}. These cultural notions of record-keeping connect early practices to today’s infrastructures of transmitting knowledge.  

As technologies evolved, so did record-keeping. Early 20th-century punch-cards supported census-taking \cite{bachman2009origin, luebke1994locating}, while digital computers in the 1950s–60s introduced new challenges in storage and access \cite{bachman1975trends, haigh2016charles, stierhoff1998history}. By the 1990s, standards emerged to ensure reliability \cite{Corbi1989IBM, johnston1997real}. In the 2000s, automation enabled attention to interpretation and quality assurance \cite{gray2002online, lord2004data, curry2010role, karasti2006enriching, witt2009constructing, baker2009data}, while the rise of \textit{big data} in the 2010s created new challenges as agencies amassed vast datasets without immediate use cases \cite{freitas2016big, pouchard2015revisiting, victorelli2020understanding, roh2019survey, chen2022merging}. HCI research intersects here not only through productive critique of these practices, but also through interactive tools for visualization, annotation, and transmission \cite{dias2020data, rezig2019towards, krishnan2016towards, tong2014crowdcleaner, greenwald2022whole, monarch2021human, van2021biological, vitale2020data, abdul2018trends, lage2011receptivity, chen2020building}. Yet, these practices often remain tied to Western-centric institutions, privileging powerful entities and marginalizing others \cite{folbre1989women, wachter2000fifth, grundmeier2015imputing, burton1994burdens, dirks2006scandal, prentice2018health, yuan2022occupational, williams2015racial, donovan2019toward, prasad2006unravelling, rycroft2006santalism, toffin2009janajati, giblin2019dismantling}. We build on this gap by examining alternative storytelling and record-keeping practices in rural Bangladesh.  

HCI scholarship of information preservation has studied the usage of interaction tools for data visualization, data cleaning, and data annotation \cite{dias2020data, rezig2019towards, krishnan2016towards, tong2014crowdcleaner, greenwald2022whole, monarch2021human, van2021biological}. Additionally, HCI researchers have contributed by designing these tools to be user-friendly and effective \cite{vitale2020data, abdul2018trends, lage2011receptivity}. The processes often involve humans-in-the-loop to interact with computers to collect data, clean data, organize data, and preserve data \cite{chen2020building, van2021biological, monarch2021human, rezig2019towards}. Note that HCI scholarship has mostly engaged with data collected and preserved through formal scientific approaches under the supervision of Western-centric agencies and entities (e.g., universities, museums, libraries, government foundations, industry, etc.). Such entities in power often risked curating biased data and manipulating data, discarding the voices of marginalized ones \cite{folbre1989women, wachter2000fifth,  grundmeier2015imputing}. Examples include historical records by colonial governments \cite{burton1994burdens, dirks2006scandal}, a collection of reports on struggles of the working class recorded by higher-ups \cite{prentice2018health, yuan2022occupational}, and racial history recorded by racial majors \cite{williams2015racial, donovan2019toward}. In many cases, such marginalized groups had limited access to mainstream communication channels, and they developed their own ways to communicate with fellow community members in contextual and situated manners, which are rarely included in mainstream histories recorded by accountable entities \cite{prasad2006unravelling, rycroft2006santalism, toffin2009janajati}. Broader debates around practices of record-keeping and collection, such as those in art, also reflect colonial trends in the appropriation of objects and the proclaimed role of colonial powers as "protectors" of marginalized communities' historical artifacts \cite{giblin2019dismantling}. Motivated by this literature, we investigate alternative forms of record-keeping and storytelling for narrative construction with marginalized rural Bangladeshi communities.

\subsection{Visual Storytelling}
Data narrative and visual storytelling are thriving areas of research that are beyond the scope of this paper. However, we will highlight a few works that have motivated our research in these areas and connect to some of the broader themes we have observed in our fieldwork. Researchers have advanced the data analytics domain by exploring how to construct visuals that transform data into visually shared stories \cite{lee2015more}. In this regard, researchers emphasized understanding the audience and setting and argued for more clarity of the underlying data and analysis processes.
Segel and Heer in particular suggested that better interactivity and usability in storytelling through visuals could be achieved by not following prescribed ordering but by integrating subjectivity in a reader-driven approach \cite{segel2010narrative}. They draw distinctions between author- and reader-constructed narratives, identifying trade-offs among the approaches. 
Willet et al. further advanced thinking in this area by creating a set of fictional superpowers that characterize how visualizations empower people by conveying information \cite{willett2021perception}. Researchers also suggested taking the storytelling visualization beyond entertainment agendas and proposed ``serious storytelling" framework for narratives of serious problems related to wellbeing, education, ethics, and religion  \cite{lugmayr2017serious}. 
Recently, Sultana et al. have explored the non-traditional visual storytelling methods in local quilts, religious idol-crafting, and witchcraft practices \cite{sultana2021seeing} as well as
how local networks of meaning provided agency in visual communication \cite{sultana2023communicating}. Motivated by work in visual storytelling, we advance this area by examining how rural Bangladeshi practices of record-keeping and visual performance organize community information, preserve local voices, and convey historical narratives.

\subsection{Factuality, Emotion, and Aesthetics in Data Narrative}
\textbf{Factuality}, which could also be termed objectivity or truthfulness in data narratives, refers to the accuracy and veracity of the information presented. Today's data science, information visualization, and data narrative research in HCI follow the positivist approach \cite{meyer2019criteria}, considering a lack of factuality synonymous with ``untruthfulness" or a ``lie factor" \cite{tufte1985visual}, and would ideally intend to minimize this factor in information visualization to sustain factuality \cite{ambrosio2014objectivity, huff1993lie}. 
It rarely values the beneficial aspects of subjectivity, which is the philosophical opposite of factuality or objectivity \cite{ambrosio2014objectivity, carpendale2017subjectivity}. However, other literature has argued that eliminating personal judgment and emotion is hard while presenting data \cite{jgrgenson1995visualization, Visualiz1:online, van2005value}, to the point that a positivist approach may not be tenable in practical settings, as
subjectivity and personal judgment in visual encoding are unavoidable in modern digital visualization \cite{van2005value}. 

Additionally, \textbf{emotion} has remained another core theme of data and narrative visualization research. Researchers working at the intersection of visualization and data narratives have argued that emotion plays a vital role in making physical embodiments of data successful in conveying intended narratives \cite{wang2019emotional}. These observations will align with some of the performance practices we outline, which seek to embody information in meaningful ways for viewers. A body of literature has also 
highlighted that emotion is key to the user experience with serious data stories \cite{lan2022negative, lugmayr2017serious, lan2021smile, bach2018narrative}. Researchers also sought to evaluate the impact of visualizations through their emotionality--- a measure of a person's emotional reactivity to a stimulus \cite{denzin1983note}. For example, one project sought to channel the seriousness of COVID-19 pandemic through negative entities in visualization so that they work as stimuli and influence people to be aware, to care, and to take protective measures \cite{COMICCop96:online, TheProje32:online, 500000do29:online}. In another example, to influence people about the Iraq War's impact, news reports presented death counts using red color upside-down bar charts in a way that would seem like blood is dripping in order to trigger visceral responses \cite{Iraqsblo67:online}. 

\textbf{Aesthetics} have remained another core factor in constructing effective visualization, physical embodiment, and data narratives \cite{cawthon2007effect}. Researchers have shown that aesthetics play a role in whether visualizations and user interfaces are usable to the users \cite{overbeeke2004beauty, petersen2004aesthetic, tractinsky1997aesthetics}. Even visuals with high-quality aesthetics might be more effective and efficient compared to similar visuals offering the same affordances \cite{cawthon2007effect, overbeeke2004beauty, petersen2004aesthetic}. Being a persuasive factor, aesthetics have played the role of incentives in a system's overall acceptance \cite{sutcliffe2001heuristic}. In this regard, researchers have experimented with bends, edge crossings, angles, orthogonality, and symmetry as primitive elements of aesthetic appeal \cite{purchase2002metrics}, all of which connect to broader art historical concepts. Projects have also characterized the specific advantages that aesthetics can provide interactive systems \cite{fogg2001makes, hartmann2006assessing, kurosu1995apparent, scha1993computationele}. Additionally, aesthetic judgment has been shown to improve the effectiveness of task performance reflected by a reduced completion time and error rate \cite{ngo2001another, kurosu1995apparent, stasko2000evaluation}. Visualization scholars have long debated whether a lack of aesthetics remains a major standing challenge for contemporary visualization design \cite{chen2005top}.
In sum, this research motivates us to examine how aesthetics, emotion, and factuality in rural Bangladeshi visual performances manifest advantages for local audiences which may otherwise be overlooked.



\section{Methods}

This paper draws on a ten-month-long ethnographic study conducted in three villages in the district of Jashore, Bangladesh. These included Rodropur, Chachra, and Shankorpur. All of these villages are located within a radius of 10 to 15 kilometers from the center of Jashore town. While Puthi, Bhandari Gaan, and Pot music are widely popular and practiced across Bangladesh, Jashore is one of the few places that make a home for all three of them as a cultural center. The ethnographer and first author was born and brought up in Jashore and is familiar with many local cultural practices. In those ten months, we studied 18 families with more than 100 members, among which 76 participated in interviews and focus group discussions. The fieldwork consisted of semi-structured interviews(n=44), focus group discussions (9 sessions, n=59), and observational field notes of rural performances accompanied by contextual inquiries and photography. This section briefly discusses the methods and relevant details.

\subsection{Access to the participants}

\begin{table}[!t]
\centering
\begin{tabular}{|rl|}
\hline
Total Participants: & 76 (Female: 41, Male: 35)\\
\hdashline
\multicolumn{2}{|c|}{\textbf{Type of Participation}} \\
\hdashline
Interview only: & 17 (Female: 10, Male: 7)\\
FGD only: & 32 (Female: 17, Male: 15)\\
Interview + FGD: & 27 (Female: 14, Male: 13)\\
\hdashline

\multicolumn{2}{|c|}{\textbf{Age range (in Years)}} \\
\hdashline
All: & 19-60, median 33\\
Female: & 19-60, median 30\\
Male: & 19-56, median 38\\
\hdashline


\multicolumn{2}{|c|}{\textbf{Education}} \\
\hdashline
No Formal Schooling: & 27 (Female: 20, Male: 7)\\
Primary School: & 15 (Female: 8, Male: 7)\\
Secondary School: & 14 (Female: 5, Male: 9)\\
Higher Secondary School: & 11 (Female: 5, Male: 6)\\
Undergraduate and Above: & 9 (Female: 3, Male: 6)\\
\hline
\end{tabular}
\caption{Demographics of interviews and FGD participants}
\vspace{-15pt}
\end{table}

Our access to participants in the villages was facilitated by the Rural Reconstruction Foundation (RRF), a non-profit global development organization that offers microfinance, education, health, and agricultural programs \cite{34RRF}. Their officials introduced us to front-line microcredit workers who visit rural clients' homes weekly. We were particularly interested in engaging with artists and enthusiasts of rural visual and performing arts. RRF workers helped the first author reach such communities by taking her to villages where they worked. After arriving in the village, the first author held a public community meeting with the microcredit clients and explained the purpose of the research study. After answering any questions and concerns the clients had, we recruited participants from the meetings based on their availability. Further recruitment was performed through snowball sampling \cite{Biernacki1981}.

The primary occupations of those families include farming, fishing, and small businesses, with an average monthly income of \newedits{approximately BDT 13500 (equivalent to USD 110)}. We developed a close relationship with those families by making frequent visits, engaging in long conversations, helping with their household and daily activities, and joining their afternoon hangouts while observing them in a participatory fashion. After building rapport this way, we slowly started joining the artists' group meetings for the performance preparation. In such meetings, the artists generally rehearsed together as well as discussed the lyrics and their explanations, choreography, costumes, stage decoration, musical instruments, and other associated artifacts to be used in the performance. In addition to the performers and group managers, local villagers also participated in these meetings and shared their feedback. While joining as observers, we recruited many of the participants from such meetings. 

The first author who conducted the fieldwork is a native Bengali speaker and has a long-term familiarity with the neighborhood. She was born and raised in Jashore and has lived in various parts of the district. This positionality helped her access the population and build rapport. All interactions with participants were conducted in Bengali, the local language, which all the participants and the ethnographer speak fluently. We obtained oral consent from the participants in consultation with a university institutional review board since many villagers were low-literate and would have trouble reading and understanding a written informed consent form. \newedits{Although some participants were more literate, the RRF staff advised us to stay consistent in ways of consent obtaining to avoid confusion, so we sought only oral consent from all the participants.}

\subsection{Observation}
We started our study by observing the villagers' performing art practices associated with cultural narrative communication. Our observation sessions spent substantial time understanding the artists' profession and working style. We observed their daily activities, hangouts, regular meetings with artist friends and colleagues, leisure activities together, collaboration on lyrics and music production, stage decoration, costume section, and choreography development for the performances. In addition, we observed their group meetings alongside the other villagers to follow how their performance was produced with community inputs. The mode of observation was participatory. We took part in each of these activities with the participants with their permission, observed their actions and responses, and recorded them in our notes. We also observed how such art and cultural practices are involved and influential in the artists' and performance enthusiasts' daily lives. We asked questions to the participants and requested an explanation about their activities if they were confusing or required domain expertise. \newedits{We conducted hundreds of hours of observation sessions with more than 100 people (including the FGD and interview participants) in the three villages.}

\subsection{Focus Group Discussion (FGD)}
We conducted nine focus group discussions with 59 participants. Each group consisted of four to eight participants. The participants of the focus group were identified through snowball sampling with the help of RRF microcredit fieldworkers. The topics included artists' careers and training, their social, economic, historical, and moral influences on the performances, and their artist-philosophy and narrative construction in performances. We were also interested in their collaborative practices within and beyond the groups. Additionally, we also asked them about different types of challenges they face in the modern day and how they resolved them. The discussion sessions were semi-structured -- the researcher led the discussion and asked related questions to engage the participants, clarified their responses, and went deeper into the topic. The questions asked by the ethnographer during sessions spurred further discussions on related topics. The sessions were generally thirty-five to forty-five minutes long. We took detailed notes and audio-recorded most of the discussions if the participants permitted.

\subsection{One-on-one Interviews}
We conducted one-on-one semi-structured interviews with 44 participants. Each interview lasted approximately 30 minutes and was conducted wherever it was convenient for the participant, often in their homes. Participants were sought based on rapport with the ethnographer in earlier sessions and further through snowball sampling. The topics included artists' careers and training, their social, economic, historical, and moral influences on the performances, and their artist-philosophy and narrative construction in performances. We were also interested in their collaborative practices within and beyond the groups. Additionally, we asked them about their challenges in the modern day and how they resolved them. We asked further related questions to understand the participants' responses and go deeper into topics. Note that our queries in focus groups and interviews were the same; we asked similar questions, letting the participants choose their methods of participation. Each interview lasted around 30-40 minutes. We took detailed notes of all the interviews, while 26 participants allowed us to audio-record their interviews (an average of 25 mins).

\subsection{Data Collection and Analysis}
We collected approximately 13 hours of audio recordings, which the researchers transcribed and translated into English. \newedits{In some cases there are no direct English words for culturally nuanced Bengali terms (e.g., Kolkata Bat-tala, Shaeri, Bonbibi, etc.). For these we relied on the closest English words and provide details while presenting findings on them.} \newedits{We took shorthand field notes, which the ethnographer later elaborated into 200 pages of detailed field notes.} We then performed thematic analysis on our transcription \cite{76boyatzis1998transforming, 75strauss1990open}, starting by reading through the transcripts, allowing codes to develop. Twenty-seven codes spontaneously developed initially. After a few iterations, we clustered related codes into themes, for example, lyrics, choreography, dance moves, materials, musical instruments, memory, history, myth, religion, homage, spirituality, etc. We present our observations in the following section.

\section{Traditional Cultural Practices in Rural Jashore}
Our field site, Jashore, was established as a district of greater Bengal in the British Empire in 1781 and is currently one of the oldest districts in Bangladesh. The British Empire constructed roads, highways, and many administrative centers to make it a political hub in the country. Consequently, this area also grew as a cultural hub in the past two centuries, especially for musicians, performing art communities, and handicraft artists. For example, people frequently organize village fairs and enjoy locally crafted and performed Puthi, Bhandari Gaan, and Pot music. Such professional performances are part of the country's cultural heritage and play a major role in national history. The Jasshorian local cultural heritage holds local historical descriptions missing in many major, organized, and mainstream recorded histories. Our ethnographic work with Jessorian Puthi singers, Bhandari singers, and Pot artists and performers studied how their performance outfits and related materials convey local stories and missing history. In this paper, we will frequently refer to ``Gazir Gaan" (Gazi in short), which is a local Bangladeshi mythological story. Bhandari Music, Pot Music, and Puthi can each have Gazi performances. 

\subsection{Case-1: Puthi Gaan}

\begin{figure*}[!t]
\centering
    \includegraphics[width=0.99\textwidth]{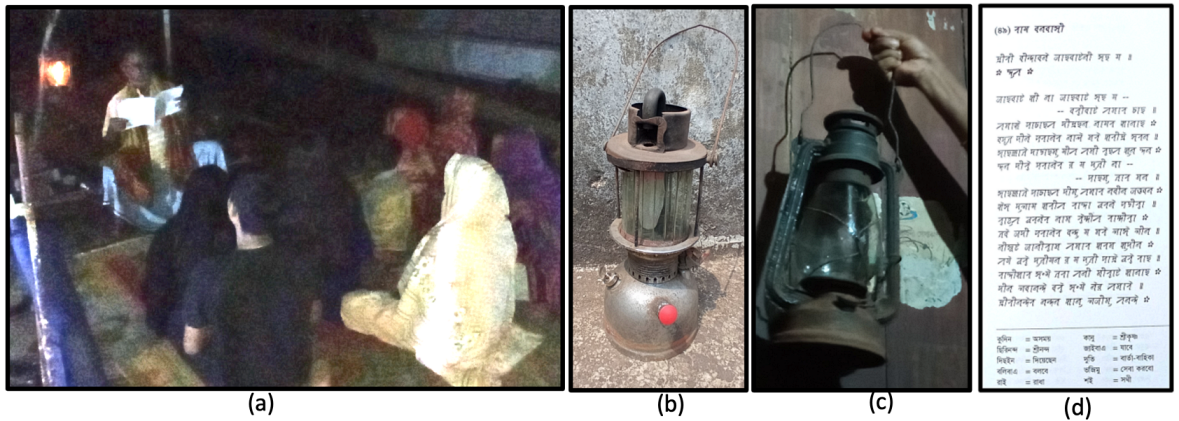}
    \vspace{-10pt}
    \caption{(a) A Puthi-Pathok is sitting on the Pati (rural sitting arrangement on the floor) and reciting the Puthi, (b) Hajak-lantern used for Puthi performance, (c) Hurricane Lantern used for Puthi performance, and (d) An example excerpt from the Puthi-book that the Puthi-Pathok was reciting. During a Puthi session, the performer would first recite the Puthi in its actual language at the top, then clarify the major word meanings following the bottom part, and then explain the takeaways.}
    \label{fig:fig1}
\end{figure*}

Puthi was a prominent form of performing arts in the greater land of Bengal during the 18th and 19th centuries and is still one of the most popular musical performances in rural areas. Puthi literature and recitals (akin to fairy tales, religious lessons, or historical stories intended for performance to audiences as oral tradition) served as the local people's suppressed voice in the early anti-colonial movement in that era, and the audiences of these performances mainly were low-paid working-class Muslims. 

\subsubsection{Performers}
Puthi performers are called Puthi Pathoks. The lyricists and composers of tones are called ``Shaer". The Puthi Pathoks were mostly low-paid working-class Muslims, including government and private employees, traders, boatmen, and peasants. While there is no gender barrier in reading Puthi, women Puthi Pathoks are very rare to find because of concerns related to harassment and communication with strangers. \newedits{For most performers, Puthi was a side gig as it pays less.}

Local people learn their Puthi recital skills from local tutors. Such training usually starts at a very early age. Puthi scripts, being a complex composition of Bengali, Arabic, Urdu, Persian, and Hindi, require the learners to master all these languages. Performers also learn the dance routines, but generally, they are encouraged not to do too much with the entire body; instead, they would want to have hand and shoulder movements. The Puthi performers we interviewed all started their training during elementary school and mastered at least two languages. They said multiple years of training are essential before they can debut. \newedits{However, Shaeri (mastering Shaer) requires more extensive training than other roles in the group.}

\subsubsection{Language, Contents, and Themes}
Puthi scripts are a composition of Bengali, Arabic, Urdu, Persian, and Hindi. The early eighteenth-century Puthi style, by Garibullah and Hamza, inherited Arabic traditional warfare history and legends. The practice then transitioned towards ``Kolkata Bat-tala" (``under the Banyan tree", meaning festivals celebrated in Hindu religious style) culture and earned the status of ``Dobhasi" (dual-lingual). Soon, this \textit{`multi-lingual poetry'} started being popular in the North-East Indian subcontinent.

The language used in Puthi often reflected both peaceful societal changes and conflicts resulting in political ramifications. After the 13th-century Turkish conquest of Bengal, Arabic-Persian words were introduced into Bengali. Following the establishment of administrative, commercial, and cultural links of Bengal with the Mughal capital, Delhi, during the 16th century, many Urdu- and Hindi-speaking people settled in various regions. Therefore, Urdu-Hindi words began to significantly influence day-to-day Bengali, along with Arabic-Persian at workplaces. This mixture of Bengali, Urdu-Hindi, and Arabic-Persian words in the Puthis is more than two centuries old. Today's Puthi uses Bengali or Nagri, with prominent colonial influence in the vocabulary.

Puthi themes can be romantic, warfare, religious, folklore, and contemporary. Contemporary Puthis talk about celebrated Muslim personalities such as Haji Shariatullah and historical events like the Wahabi-Faraezi movement. Jalalatul Fokre describes the hostility of the Orthodox Muslims towards the Baul community. However, Puthis that describe such contemporary events are rare. Describing old history, the Puthi Pathoks create a world of fable and heroism away from reality for the enjoyment of audiences.

\subsubsection{Performances}
Puthi recitals can be both solo and group performances. In the old days, Puthi performers lit large kerosene Hazack lamps and recited Puthi under an old village tree during festivals and as a major source of entertainment to villagers. They were also performed on boats resting near banks to celebrate successful business trips by traveling merchants. The choice of location, ambiance, and stage decoration is often influenced by the themes and messages the Pathok wanted to leave in the community's memory. Many merchants also used to hire their own Puthi groups for the entire trip. Their additional job was to bind memories in Puthi music that could be recalled in the oral tradition. Today, Puthi performers still light Hazack lamps when performing at village fairs, but they also now use modern hurricane lanterns as a similar symbol of heritage. These lamps are significant in signaling the mood, ambiance, and context of Puthi performances.

\subsection{Case-2: Bhandari Gaan}

\begin{figure*}[!t]
\centering
    \includegraphics[width=0.99\textwidth]{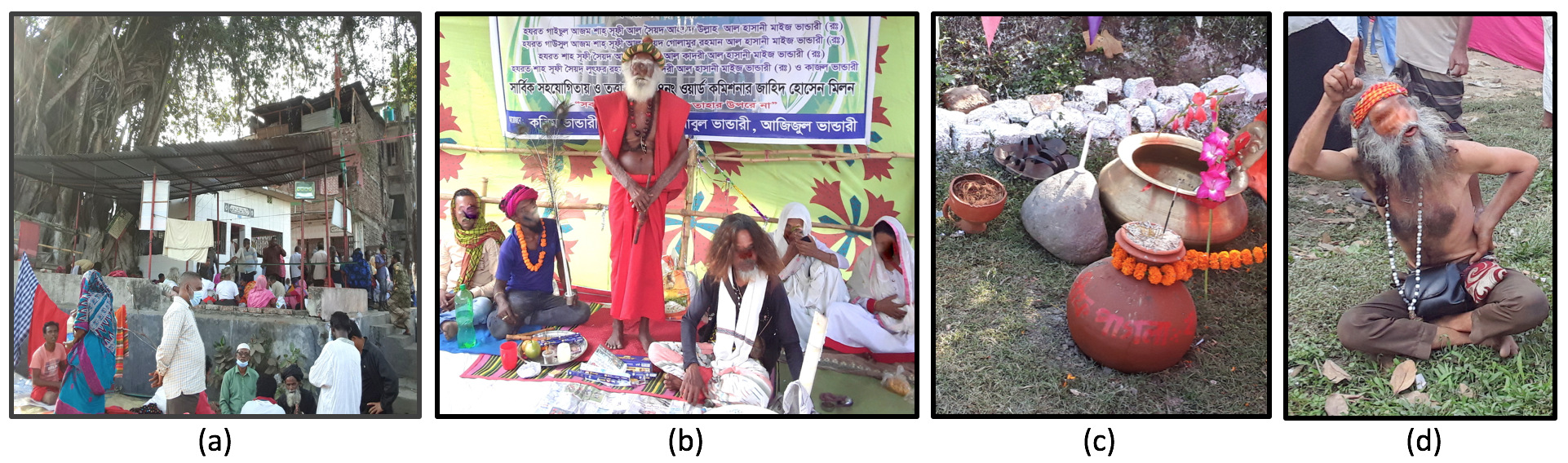}
    \vspace{-10pt}
    \caption{(a) The rural Bhandari music festival is taking place under a locally renowned Banyan tree, where people from different religions have gathered to enjoy the show, (b) One of the Bhandari teams -- the person in the red outfit in the middle is the leader. Such groups frequently include women performers, (c) Agar Batti (an incense that releases fragrant smoke when burnt) in clay pots in preparation for the performance, and (d) A performer showing off his outfit and explaining the symbolic meanings of the components.}
    \label{fig:fig2}
\end{figure*}
The word ``Bhandari" was named after the renowned Bengali Sufi \jrz{saint} Ahmad Ullah Maizbhandari. He founded Tariqa-e-Maizbhandaria, a genre of Islamic lifestyle that adopted instructions from Maizbhandari himself. The music produced based on the lyrics written on Maizbhandaria's history, scholarship, and tales is named Bhandari Gaan (the music of Bhandari). \jrz{While the origins of the music trace back to a Muslim historical figure, performances are often attended by individuals of all faiths.}

Sufi Ahmad Ullah Maizbhandari established his institute, namely ``Maizbhandari Astana," and grew his scholarship in the Bangladeshi western coast. Alongside being a lawyer in Jashore, he taught in many Islamic schools and Madrasas in the district and lectured on his divergent philosophical standing regarding contemporary politics, society, and religion. Based on his lecture, the musical genre has evolved and remained immensely popular for several centuries now. Bhandari Gaan and lecture, having significant Muslim influence, are locally considered both a form of ethical education and entertainment.

\subsubsection{The Performers}
Participants reported years-long training. Three participants mentioned that they spent months in the Maizbhandari Astana in Chattogram to learn the new vision of spirituality. This step is required prior to musical skill development. The participants reminisced that in the early 1980s, they locally formed a 30-person group and traveled to Bhandari Astana for training. Several months later, they returned and trained local disciple groups. 

Bhandari artists generally form a company of 15-20 people. They welcome both male and female artists in their groups as long as they are well-trained and qualified to perform. The performers are of two types: vocalists and instrumentalists. Some members master both singing and instruments. A strong, steady, melodious, and loud voice is a must for vocalists because performances may take place in busy settings without electronic speakers and amplifiers. \newedits{Both vocalists and instrumentalists are required to master choreography consisting of specific styles of gesture and dance routines. Bhandari performance is seasonal, so all the performers told us this was a part-time job for them.}

\subsubsection{Themes, Contents, and Instruments}
Bhandari music often mixes Islamic stories with local flavor. Their stories claim to have happened with or been experienced by Maizbhandari. Many of those songs were written by Maizbhandari or by his contemporaries. Performers might also write their lyrics and send them to the chronicle. The Bhandari artists we spoke to produced over fifty songs for their repertoire, but they never wrote the lyrics down anywhere because they lacked literacy. Maizbhandari scholars set strict rules on the topics and themes of the lyrics and instrument use. Generally, the performances use Dhol (two-headed drum), Tabla (one-headed drum), Jhunjhuni (bells), Ektara (drone lute), Dotara (plucked string instrument), and Harmonium (reed organ). 

\subsubsection{Performance}
Bhandari groups perform seasonally, during religious festivals and village fairs. They also perform at conferences and festivals organized by other religious followers on request. Generally, performances are open-air or under the village Banyan tree. They specifically chose the Banyan tree because this species holds value in villagers' spiritual sentiments. In an ideal setup, Bhandari tents and stages would be side-by-side for easier movement of performers and instruments.

Bhandari costumes signify holy connotations. Generally, leaders wear bright-colored attire. Local religious and spiritual values influence the choice of stone and accessories. The male performers also wear caps, different from those used in Muslim prayers. The female performers wear a sari, preferably of white color. They use Surma (local eye makeup) but are encouraged to be otherwise sparing (though they can put on stones and ornaments). Burning Agarbati (incense used by Muslims) and Dhup (incense used by Hindus)  before a tent indicates that the performance will start soon. They also offer free food and snacks during the performance and accept donations. The performances generally start in the evening and may run all night. Such festivals might be one day long but may also last a week.

\subsection{Case-3: Pot Tales}

\begin{figure*}[!t]
\centering
    \includegraphics[width=0.99\textwidth]{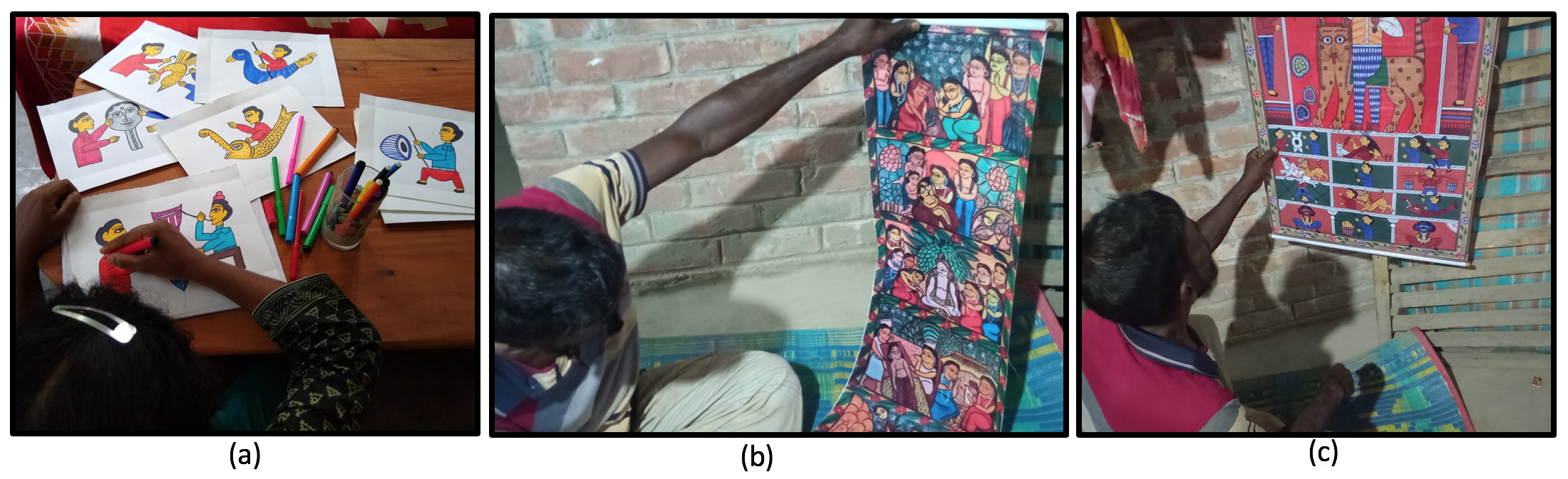}
    \vspace{-10pt}
    \caption{(a) The process of preparing a Pot tale scroll polyptych, the person in the image is drawing the chunks first to glue them together on a scroll later, (b) an example scroll after the sections are glued together, and (c) the Pot tale performer is explaining each chunk of the Pot to the ethnographer through a sample performance.}
    \label{fig:fig3}
    \vspace{-10pt}
\end{figure*}

Patachitra (Pata or "pot" [scroll/canvas] + chitra [images]) is a 2500-year-old ancient folk art of greater Bengal, including West Bengal and Bangladesh. Synonymously known as Pata, Patua, Poter Gaan, Patachitra, and Patua Sangeet, this cultural heritage plays a significant role in the Bengali language and traditions. We adopt the informal common term "pot tales". The performer, termed as \textit{`Patua'} shows a set of paintings of a historical event or folktales and interprets them to the audience through song and dance. This could be a solo or group performance. The paintings are generally combined onto one polyptych in the form of a scroll or canvas of connected scenes. It is particularly famous in the rural parts of the country, including the villages in Jashore. Patachitra is particularly valued by Bengali art and culture enthusiasts for its brilliant play of color that combines to showcase many century-old significant events and experiences in this region. Bengali historians also argue that Patachitra is the world's first attempt to create motion pictures \cite{mondal2022discourse}. For community members with low literacy, Patachitra is an engaging way to learn and pass on folklore, moral lessons, and historical drama.


\subsubsection{Performers}
Traditionally, Patuas would have been trained for years. \newedits{They would train in painting, singing, and dancing. However, most of today's Patuas work as a group, with some members painting and others mastering vocals and dance routines.} Our participants informed us that previously, there were specific Patua villages where they lived as a community and trained disciples. These communities have died out over time due to pressures from radio, television, and other more monetizable forms of entertainment, leaving Pot on the verge of extinction. Patuas are generally at the bottom of the rural economy. Many of today's Patuas are part-time performers and full-time farmers, as well as rickshaw and van pullers, easybike drivers, and factory workers. To keep up with the times, many Patua painters also decorate handicrafts, household items, and textiles, selling them to promote sustainable living. Some Patuas showed the ethnographer scrolls that illustrated their current lifestyle as part-time Patuas.

\subsubsection{Themes and Materials}
Traditionally, Patuas write their lyrics and paint their scrolls. First, they use paper or bright-colored cloths to draw the frames or unit paintings that explain the subsection of the story. They use vibrant colors derived from natural products, such as vegetables and turmeric, to create the frames or paintings. Once all the units are made, they place the unit frames sequentially on a longer scroll of cloth or Jute-mats to form a polyptych. The scrolls are three to fifteen feet wide and two to eight feet or more in length. However, many of today's Patuas borrow from other Patuas' scrolls or locally available scrolls for their performances due to resources and time constraints.  

\subsubsection{Performances}
Patuas perform by singing tales and collect donations by moving from door to door. They also used their skills to make clothes and household goods containing scroll scenes. The economic constraints have pushed these performances to become seasonal. Therefore, Pot artists are rarely found performing in villages unless they are invited to festivals or fairs. Sometimes, the local government subsidizes Pot programs for national initiatives or to promote awareness in the villages where Patuas use their specialized Pots. Recently, the government's initiatives to promote forestation, environmental sustainability, women's health awareness, and many other important movements utilize Pot tales, leading to promoting both the art and specific agendas in publicly funded performances.

\section{Findings}

\subsection{Embodying Narratives through Performance}

Participants told us that they consider the performances of Puthi, Pot, and Bhandari Gaan as community sources of information which are curated to achieve specific narrative, emotional, and religious aims. \jrz{While there are common framings provided by the \newedits{\textbf{material and oral culture}} (such as a Puthi story)}, performers have flexibility in determining the individual aspects of their performance. Different attributes, such as their outfits, gestures, and body language, play important roles in narrative-making while leaving their audience with a lasting impression.  
For example, the Bhandari performers' outfit and styling convey significance and even specific characterizations. The performers put on red or bright-colored clothes with lots of jewelry if they were playing positive-protagonist characters, or might instead put on dim colors if they portrayed negative attitudes to provide non-verbal signals of the intended mood. \newedits{The excerpt from P22 explains the \textbf{local visual conventions}},

\begin {quote}
\textit{``Gazi will wear red and Kalu will wear blue. Also, no other character should wear a red outfit at all. This is how we prepare the pots, so people will know who and where to pay more attention.", (P22, female, 29 years)}
\end{quote}

Five participants discussed \newedits{\textbf{visual conventions}} of how the artists portray time-specific information in their performances. They told us that to portray the verse of a character from the past, most Gazi performers would change to a faded fabric outfit, and the main character would put on an older-fashioned hat. In Pots, the scroll would convey continuity information in the borders and positioning of the pots, as in many other forms of polyptych art such as altarpieces in northern Europe. Unlike these other forms, however, borders offered much more \textbf{narrative flexibility} and did not follow strict conventions (as in altarpieces, which generally follow text reading order or continuous narratives). For example, a Pot artist showed us the borders of the pot (see Fig.4) and explained how their variations denoted sequence, consequence, parallel sequences, and ending.

\begin{figure*}[!t]
\centering
    \includegraphics[width=0.75\textwidth]{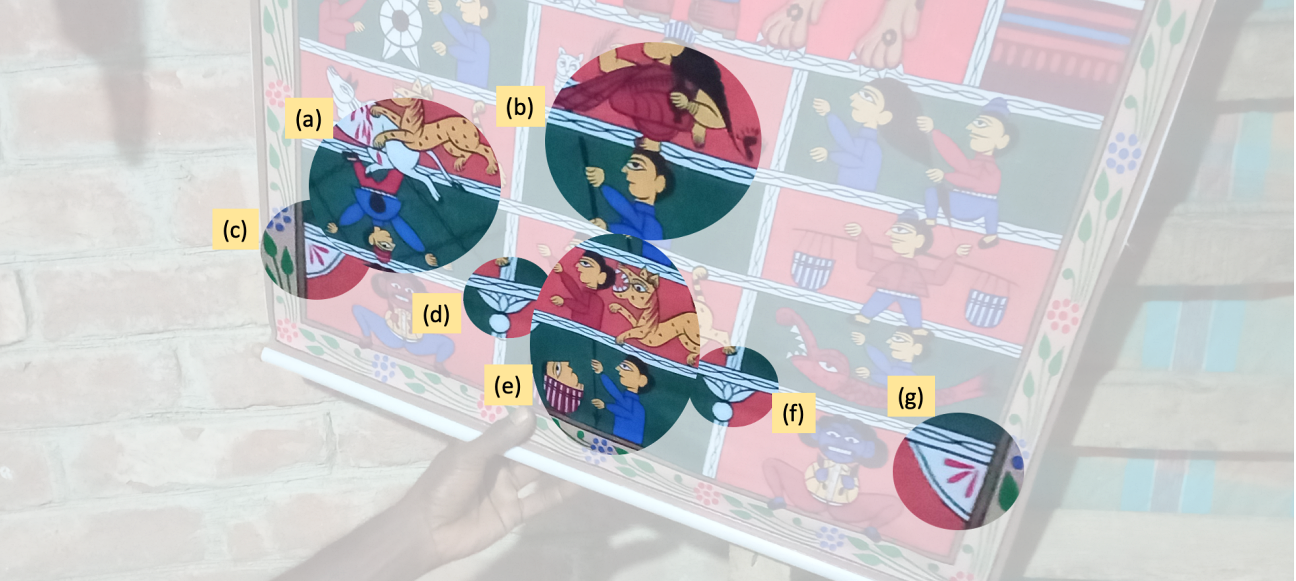}
    \vspace{-10pt}
    \caption{(a) When the tiger attacked the person in the blue shirt, he was able to escape as it shows he could cross the border of that sub-pot, (b) the person in the blue shirt on the bottom sub-pot had Goddess Laxmi's blessings with him from the top sub-pot, as the foot of the Goddess crossed the border of the sub-pot, (c) and (d) floral marks with specific meanings, (e) the person in red shirt on the top sub-pot could not escape the tiger's attack, and person in the blue-shirt on the bottom sub-pot is collecting his head, and (f) and (g) floral marks with specific meanings.}
    \label{fig:fig4}
    \vspace{-5pt}
\end{figure*}

\begin {quote}
\textit{``See (pointing at Fig.4(b), his wife's prayers were strong enough that she is portrayed as Goddess Laxmi, and the foot of Laxmi is crossing the border as it was strong. Now see here (pointing at Fig.4(a), when the tiger attacked them, Gazi could escape by crossing the border to a different dimension in time. But Kalu could not escape (pointing to Fig.4(e)), as Gazi collected his head afterward. It happened afterward, as the tiger's foot crossed the border. The petals of this type (Fig.4(d,f) denote takeaway or consequences, while petals of this type (Fig.4(c,g) denote the end of the story, and the Karma served everyone.", (P39, Male, 41 years)}
\end{quote}


The participants also reported that they frequently used specific gestures and postures and sometimes symbolic dance moves, which were either related to specific gestures common in the culture or conventions of the performance art (reminiscent of traditional practices in \textit{pantomime} and \textit{rakugo} storytelling, for example). This also allowed the performers to convey messages that otherwise would be hard to convey in dialogue or because of their social status. For example, a male performer explained what movements he would use when he depicts Gazi's wife,

\begin {quote}
\textit{``She was more like a devoted wife. Such characters are compared to the Hindu Goddess Laxmi, even if the women in question are Muslim. To portray Gazi's wife, I would make a sign of an owl that always accompanies Goddess Laxmi, and the sign of Laxmi's feet using my hands so that people will understand the woman is a calm and nice lady, even when I would only say "the wife".", (P39, Male, 41 years)}
\end{quote}

The Pot performers informed us that they would add more hip movements to portray sexually indecent behavior. They explained that talking about sexual harassment is a stigma in rural communities, and they considered that such portrayals might work as an indicator to teach people what gestures, postures, and behaviors are vulgar and unacceptable. If an adversarial character models a behavior, it can imply that the behavior is undesirable. Such ``undesired" come from the society's common, shared understanding, and the performers infuse them in their choreography to preserve and communicate such knowledge across generations. \jrz{Performance creates a dialogue between the performer and audience member where cultural conventions are both taught and referenced.} P17 explained,

\begin {quote}
\textit{``The bad guys make nasty movements. So when you see them doing bad things, you will immediately know they are the bad guys. Even after the show, people will remember and reference those gestures and postures as indecent, and some might even learn and teach others the "dos" and "don'ts".", (P17, male, 29 years)}
\end{quote}

Similarly, Puthi Pathoks also acquire and preserve myth-, folktale-, and spirituality-based moral knowledge sets by choreographing \newedits{\textbf{(i.e., embodying)}} them into memorable gestures and postures for audiences. A Puthi Pathok told us how he improvised his recitals involving animal cruelty in the middle of the performance, so that the audience would pay extra attention and remember the hidden urge for empathy.

\begin {quote}
\textit{``...(I)n one of the most popular Puthi, a deer is hunted at a tender age. Before getting killed, it cries and begs to be set free and go back to its mother. So, in the middle of my performance, I can not stop and lecture people on animal cruelty, right? So, what I do is make my voice extra loud when I read about the deer's screaming and crying part. I also play those verses twice, even though the instructions say I should be doing it once. I do it so that even children who came to see the performance remember my scream for a long time after the show and empathize with the crying deer.", (P19, male, 42 years)}
\end{quote}

Thus, Puthi Pathok manipulated his vocal tone, scale, and volume to provide non-verbal signals that reinforced his lessons more intensely than a lecture could, even though it broke with the official canon. Twenty-three participants said they often recall when a Puthi Pathok cried or shouted out, so those are significant points of interest.

Another 15 participants told us that their local Puthi Pathok often practiced with selected villagers and sought feedback from them on their tones and scales to calibrate for better communicability and memorability. Thus, the participants explained how different vocal aspects in the performance were influential in curating the narratives. \jrz{In other words, performers placed emphasis on non-narrative aspects of their performance, which they viewed as more important and necessary to remember than the narrative itself.}

\newedits{Taken together, the bodies, materials, and gestures in performances become integral narrative instruments, allowing stories to be felt as much as understood. Through movement, color, voice, and shared symbolic cues, performers and audiences co-create meanings beyond spoken text, sustaining moral lessons, and cultural memory through embodied expression.}

\subsection{Blending Myth and History in Favor of Narrative Goals}

The three performing arts all frequently used examples that straddled the boundary between fables and history. While they often contained historical figures or purported to reflect a real timeline of events, they would be hard or indeed impossible to provide in a traditional historical context. For example, the artists used stories from Pirs and Awlias who existed in myth, and the names of the places, stories, and lessons included in those stories rarely existed in the real world. The communities we worked with did not focus on historiography; rather, the lessons and takeaways were important to them. They would hardly question if such a place ever existed on the geography or maps, or if recorded history would be able to confirm if any such people were there in the given timeline in the stories. \jrz{ This aspect is shared with many other parables, fables, or fictions internationally. In the case of the performing arts we observed, however, an extraordinary amount of attention was paid towards situating the setting and narrative structure in ways that would be meaningful to local audiences and local history.} 
For example, when we asked one of the Bandari performers and songwriters about Bhandari's actual identity and timeline (given copious records indicate he did exist as a historical figure), he said, 

\begin {quote}
\textit{``Why do you care about his timeline? I know my father knew his stories, and my grandfather also used to tell his tales. My Ustad (master teacher) said Bhandari Shaheb was known to his community and other known communities for more than five generations. It does not matter which year Bhandari Shaheb was born. It does not require any proof for us to inhale his lessons.", (P37, Male, 55 years)}
\end{quote}

However, the participant later informed us that the Sufi was believed to have been born sometime in the early 1800s and started teaching Sufi lessons at a young age. For a long time, Bhandari Shaheb instructed his fellows about religion and spirituality and, therefore, started his own Bhandari culture. Performers were willing to fluidly shift between historical fact and fable in order to serve their \textbf{narrative goals}, and showed comfort in \textbf{maintaining both perspectives simultaneously} in a performance. \jrz{Audiences, in turn, accepted even potentially contradictory events in favor of the intention of the performer.} 

Similarly, the sequentiality, succession, and time length of events in these myths and stories were less important to Pot musicians. Instead, they emphasized the narratives and messages to convey in their performances. Hence, these were crucial in their construct of factuality. In a discussion where we talked about Gazir Gaan in Pot music, different performers gave us different timelines of Gazi's life events. When we questioned them about this, they referred to other participants' information,

\begin {quote}
\textit{``Do not worry if I say Gazi and Kalu prayed in the jungle for seven years and someone else says it was seventeen years. Neither of us has evidence. Interestingly, it does not matter! What matters is how they used the long-term prayer in constructing the lessons here. As the story says, their long prayers went in vain because they disheartened their mother and disregarded her opinion --- seven or seventeen --- their prayers were useless.", (P21, Male, 61 years)}
\end{quote}

Thus, the Puthi, Bhandari, and Gazir Gaan performers pointed out that the \textbf{ethics and morals} inherent in the performances take precedence over historical accuracy. While these performance traditions are hardly the only ones globally that \textbf{prioritize narrative and message} over causal rigor, our participants were notable with respect to the comfort that they showed in accepting simultaneous (and sometimes contradictory) truths. In many cases, evidence came from the performative effect and not the specifically attributable and objective causes of an event on a pot scroll or a performance of Gazi's life. Also notable was the way that performers found solidarity in each other's variations of stories and demonstrated comfort with contradictions among already contradictory depictions of historical events. Instead, local ethics, subjective lessons, and takeaways were prioritized by performers because they did not find that objective truth was a core value in how they curated their message.

\newedits{Taken together, these performances show how communities fluidly blend fable, memory, and moral judgment to shape narratives that are collectively meaningful even when historically inconsistent. Rather than privileging objective accuracy, performers and audiences co-create stories that preserve ethical lessons, affirm cultural identity, and maintain continuity across generations despite contradictory details.}


\subsection{Visual Aesthetics and Shared Social Context as a Resource}
We observed that the performers highly emphasized visual aesthetics for effective communicability, memory-making, and future reuse and sharing of narratives. We documented a variety of aesthetic techniques used in Puthi, Pot, and Bhandari performances. For example, in Pot music shows, the Patuas wanted their scrolls to be visually high quality with bright colors and smooth yet exotic shapes so that the audience would pay attention and it would hold their interest for longer periods of time during the performance, and onward. P37 explained,

\begin {quote}
\textit{``With good quality images on the pot, the audiences get hooked to them. They remember the details more and want to learn more. Such smooth-looking details give the story life, so the audience pays attention. If my pot characters wear colorful outfits, they are more eye-catching." (P37, Male, 55 years)}
\end{quote}

Fifteen villagers told us that they often recall characters on Pot from their outfits' colors and sizes in the pot-images. The conventions that we alluded to in a previous section for portraying characters were indeed well-recognized by viewers. For example, characters in red and green would be generally the protagonists, and characters in darker shades or black outfits would be the antagonists. 

In Gazir Gaan in Bhandari performances, the performer would keep emphasizing the colors of the characters' outfits, their heights, and related descriptions to remind the audience of the \textbf{aesthetic attributes}. P34 took the ethnographer to show her around the preparation area of an upcoming village fair. We quote her,

\begin {quote}
\textit{``(The performer in Fig.5(b)) has two necklaces: one with Radha-Krishna and another, maybe a Tabiz (with Muslim scripture). This indicates he is playing a character who delivers social solidarity dialogues. Both the symbols on his body mean he respects both. The other guy (performer in Fig.5(c)) has many different colors of gemstone jewelry. Every color has different meanings. For example, he wears the green ring because he has been long-trained at the Bhandari school. He also wears the orange stone ring, which means he has been performing for a long time and is a top performer now; therefore, he conducts closing remarks." (P34, Female, 32 years)}
\end{quote}

\begin{figure*}[!t]
\centering
    \includegraphics[width=0.99\textwidth]{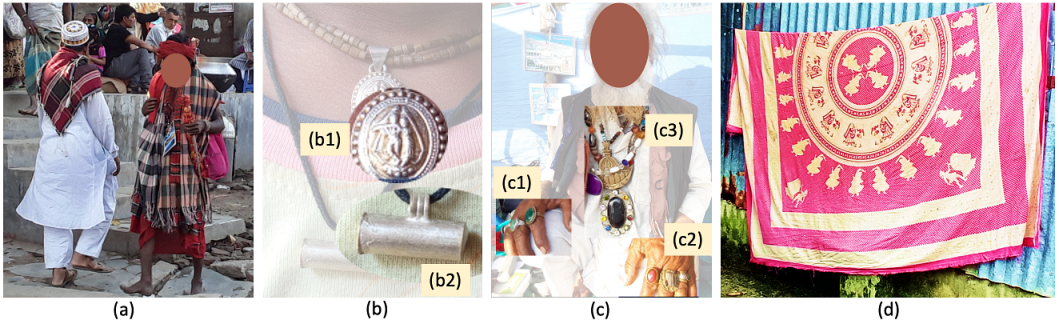}
    \vspace{-10pt}
    \caption{(a) Two Bhandari performers wore outfits that signified their narratives in their upcoming performances, (b) A performer wearing necklaces indicating (b1) God Krishna and Goddess Radha, and (b2) an amulet, (c) another performer wore jewelry of different colors in their fingers (c1 and c2) and neck (c3), (d) Hindu religious scriptures and symbols on a curtain that is going to be used as a background of performance for spiritual ambiance.}
    \label{fig:fig5}
    \vspace{-5pt}
\end{figure*}

The participants also informed us that often the size of different artifacts and the perceived sizes of the characters play crucial roles in performances. For example, the vital characters in each frame of the pots are portrayed in the biggest sizes, while the exact same characters might be smaller in size in a second frame if they are not vital in that frame. This technique, called \textbf{hierarchical scale} in art history, is a long-lived visual tradition in the region. Though it breaks with objective reality, the skewed depictions again help to further the narrative aims of the piece. In Bhandari performances, the characters make their entry and exit by curling up their body so that they are smaller in size and their presence in the frame would not seem sudden to the audience. However, Gazi Gaan employs different entry and exit strategies. P32 explained, 

\begin {quote}
\textit{``...(W)hen it is time for a new character to enter, they would slide in diagonally, unless it is a protagonist. If it is a king, others would slide in vertically or horizontally onto the stage, and he immediately stands up and starts his part, while antagonists would already start delivering their dialogue right outside or at the edge of the stage, and once the audience notices them, they will enter. And then upon finishing their part, they will leave like wind, while the king will slowly head out, and everybody will be bowing.", (P32, female, 39 years)}
\end{quote}

In a film or staged performance that aims for \textbf{realism}, such character entrances and exits would break audience immersion due to their unnatural movements. However, the participants explained how their performances employed shape, angles, and symmetry to contribute to the visual aesthetic and, hence, narratives. For example, three Patuas told us that female performers with makeup to seem round-shaped faces would represent joy, while their makeup to portray thin faces would represent adversity. Similarly, entities with a limited amount of bending would represent a positive vibe, while frequent leaning and hip movements would represent brazenness, as we recall about Gazi Gaan. 

While discussing Pot music, villagers told us that channeling balance and harmony through visual cues is highly expected in Pot performances. Eleven villagers told us that they could understand the plot and catch up with the scene of a particular performance just by analyzing the symmetric and harmonious information of the characters' outfits and the frame, as per P23, 

\begin {quote}
\textit{``... (E)ven if I missed the first few minutes, I would look at the outfits. If it is red and has a lot of symmetries of patterns and motifs on it with golden thread-works, I would know it must be either Shah Jahan's or Akbar's tale or some other Mughal lord's tale. Also, their chairs will have Lions on both armrests. You will know they are not some fake lords from some other time; this is very specific.", (P23, Male, 55 years)}
\end{quote}

Furthermore, five FGD participants showed us the curtain on Fig.5(d) and explained that such clothes with symbols are crucial in priming the audience's mood and creating the environment for specific performances. In addition to visual aesthetic components, the rural performers also played with the tones and volume of their voices and used different musical instruments to create ambiance. Such an ambiance was significant to communicate with the audience's emotionality, especially for the performances with awareness lessons and moral education. 

\newedits{As a whole, visual forms of these performances becomes a social language through which characters, moods, and moral positions are conveyed with precision and subtlety. By drawing on shared conventions of color, proportion, gesture, and symbolic adornment, performers craft culturally situated narratives that audiences intuitively grasp and carry forward, while leaving scope for polyvocal interpretations and stories that evolve across generations.}

\subsection{Moral Narrative Construction with Contextuality, Situatedness, and Tradition} 
Our participants also informed us that moral narrative construction was crucial in rural artistic performances. For centuries, these performances have been playing the role of moral learning sessions in rural Bangladesh. The morals in these artistic performances are highly drawn from local myths, religious and spiritual sentiments, and historical events (a previous subsection also briefly noted). Therefore, such local-practices-based moral standards rely highly on contextuality, situatedness, and traditions. 

For example, while secular ethics mostly discards drawing on beliefs in supernatural revelation or guidance and solely focuses on human faculties such as logic, empathy, reason, and moral intuition, ethical lessons inherent in rural performing arts often freely \textbf{mix logic, empathy, reason, and moral intuition with lessons drawing on supernatural and spiritual beliefs}. Such supernatural and spiritual beliefs are widespread yet context-specific and dependent on the history and myths of those communities. We noted that the rural performers emphasized conveying ethical lessons over the truthfulness and believability of the stories, even though they may concern historical figures or events. For example, P37 recommended we focus more on the inherent moral lessons and less on the legitimacy of the event, while explaining the myth of Gazi and Kalu making amends with 900,000 tigers upon being attacked, 

\begin {quote}
\textit{``Gazi and Kalu promised the tigers that they would not harm the habitat and requested the tigers not harm them either, and let them stay in the forest. The tigers agreed and became Gazi's subjects, leaving them alone to pray. Now, you logical modern people can not relate because you might ask how a human can convince and make promises with tigers. I have seen people laughing, saying how come 900,000 tigers agreed to that point and saw them asking if tigers had voted and practiced democracy and all those sorts of crap-talk. If you focus on the logic of the process, you are missing the entire point of the promises --- do not harm the forest and the animals, and they will not harm you in return. This lesson does not require any proof. It just has to be done.", (P37, Male, 55 years)}
\end{quote}

The participants also explained that this verse of Gazi's Gaan is very important to the people who go to Sundarbans (the forest next to the areas) to extract wild honey and fish. They recite this verse as a promise to the forest that they will not harm the wildlife and help preserve the forest's ecology. In return, they expect no harm from the forest. However, the villagers also opined that many people do unethical and illegal business on boats inside the forest, and therefore, such actions harm the forest's ecology. The forest gets angry and turns its back on the villagers. P33 explained, 

\begin {quote}
\textit{``My husband is a part-time farmer, and he also collects honey in the Sundarbans. Lately, the forest is not the same because people who are less familiar with the forest go there and kill Fishing cats as soon as they see them, regardless of the fact that the cats are not threatening them. They do not know that such cats will not attack people unprovoked. People who live near the forest have known this for generations, but new people do not know. They all come with modern guns; they say they are good hunters and have read a lot about the forest. But what's the point of being modern with training, readings, guns, and machines if you understand even the basics that even we, the uneducated villagers, know from folk songs!", (P33, female, 38 years)}
\end{quote}

The villagers were angry that many people took pride in hunting wild animals. They also discussed religious stories where fellows of Muhammad (the Muslim Prophet) fought tigers and killed them, and how many people interpreted it as a sign of heroism and proof of that person's strength. Thus, our participants explained to us how their understanding of forest and wildlife preservation might have possibly differed from the logic and recorded religious histories of hunters. Instead, their understanding of myths and stories cohered with their experiences, and therefore, they relied on the ethical lessons from the rural performances regardless of veracity. 

While discussing human greed and the environment, the participants explained how their communities have always followed the instructions in traditional musical performances that talked about dos and don'ts regarding the community's harmony. For example, an FGD participant told us about a performer group that has been active in her village for decades, conveying such messages,

\begin {quote}
\textit{``Since my childhood, I have been seeing Hasmot uncle's (pseudonym) group performing Bonbibi's Tale (means the tale of Sundarban Goddess) and talking about risks of fishing in the deep forest during the months of Asharh and Shraban, because that is the time Bonbibi protests her infants, like fishes and bees. Our parents' village has been fishing for generations; traditionally, they have avoided fishing in the deep forest in these months. Nowadays, the government notice on the TV says you can not fish during June, July, and August because that is harmful to the country's fish breeding.", (P44, female, 26 years)}
\end{quote}

Note that here, the performers were using their age-old situated knowledge to request fishermen to avoid fishing in particular Bengali months in the monsoon, which intersected with months in the English calendar in a government notice. In a follow-up question, we asked the group about the source of information used in that performance. P46 argued that this was situated knowledge, and has traditionally been followed along. She explained that her parents' neighborhood and the community in a different village also followed a similar mythological tale, where they emphasized restricting entry to the forest in certain months, 

\begin {quote}
\textit{``Before, there was no TV or much radio, so jobless fishermen and farmers would seasonally form performing groups to entertain the village. They have had experiences for ages, and their knowledge from experiences is passed down to the generations. These restrictions are renowned: whenever people become greedy and try to supersede Bonbibi, something wrong happens. So these are big NOs.", (P46, female, 49 years)}
\end{quote}

It is noteworthy that many of today's environmental sustainability campaigns subscribe to similar messages to persuade people not to harm the forests and wildlife so that we all can coexist in the ecosystem, and here, rural performers are navigating this persuasion through their contextual and traditionally situated moral narratives without any formal training on environmental sustainability.

\subsection{Adaptation to Modernity}
We also noted how the rural performers adapted to the modern needs of their audiences. For example, two Gazi musicians were preparing special songs with subunits of their musician groups for upcoming political events in their areas. They explained that while those songs were not exactly what would be called Gazi music, they were preparing those following the theme, tone, and mood of Gazi music's exclusive essence. Other subunits of their groups were also preparing the performances for the events, as they mentioned, 

\begin {quote}
\textit{``I took a contract to prepare a show for the political campaign of the local chairman election, and I am preparing the music composition that includes his contributions to the village. My colleagues are preparing the choreography. This is a campaign; we are going to give a lot of positive vibes. We will wear peaceful colors, like green and white, to convey that this candidate has always remained peaceful in the neighborhood. We have to explain his good deeds through sober movements, and we will bring our domestic doves, so people are persuaded. These are not exactly Gazi music, but inspired.", (P43, Male, 49 years)}
\end{quote}

Similarly, some Pot musicians said they used many concurrent incidents and concerns in their arts and performances. For example, during the time we were conducting the ethnography, individuals were universally concerned with COVID-19. The villagers showed us their pots on COVID-19 and explained,

\begin {quote}
\textit{``We portrayed the danger of the virus using scary bodies; we colored them in black and orange and made blood stains on them. They have sharp nails and spikes on their body. The singer would use deeper vocal scales and low registers to make it look very serious while performing these Pots. And the others would sing for the surroundings would sing in high registers and in a tone that they are scared. This performance will have a lot of sudden screaming and also shivering.", (P49, Male, 41 years)}
\end{quote}

The participants also showed us their pots related to an AIDS campaign (Fig.6(a) and (b)), in which they collaborated with local hospitals and NGOs to raise awareness about the disease. We observed their pot-making sessions, and we found out that they emphasized unsafe polygamy as a reason for HIV contagion. We discovered that they portrayed mythological demonic characters and used spiritual sentiment to grow awareness (Fig.6(b)). Rural villagers also showed us their pots about tree plantation and green tree-looking outfits and explained to us that every year, they participate in tree-planting campaigns in the months of June and July to raise public awareness about the harmful effects of deforestation and motivate them to plant more trees. 

\begin{figure*}[!t]
\centering
    \includegraphics[width=0.8\textwidth]{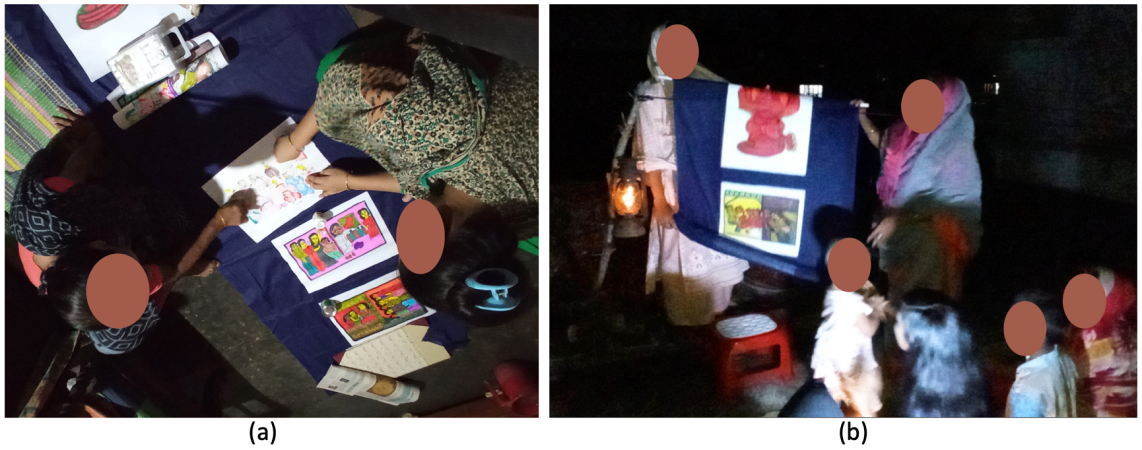}
    \vspace{-10pt}
    \caption{Rural artists were drawing images to create pots for HIV awareness performance, where (a) they emphasized unsafe polygamy to be a main reason for HIV contagion, and (b) an image taken during the performance where the performers were showing the pots contained demonic characters from local myths and signing along to worry the audience}
    \label{fig:fig6}
    \vspace{-5pt}
\end{figure*}

All the musicians also told us about the technology they use to find help in their profession. Many of them mentioned using mobile phones to access YouTube and watch other similar musicians' performances. Nine musicians were active users of social media, and they were members of Facebook-based musician unions, where they often discuss upcoming village fairs and folk music conferences, performance deals and contracts, the minimum wage for performances, and so on. Additionally, seven musicians often prepare short reels and videos of their performances and practice sessions and post them on TikTok and Likee to advertise their upcoming events. 
 
\section{Discussion}
This paper presents our findings from a ten-month ethnographic study that investigated how these communities utilize entertainment and cultural practices, namely \textit{Puthi}, \textit{Bhandari Gaan}, and \textit{Pot music}, as means to curate information and communicate traditional moral lessons and history. \newedits{Our fieldwork revealed that these communities embrace polyvocality and multiple ethical frameworks in their performances (see 5.2), construct narratives that combine factuality, emotionality, and aesthetics (see 5.3, and 5.5), and adapt their performances to changing technology and audience needs (see 5.5)}. \newedits{Our findings inform the HCI design theories and methodologies of data curation, storytelling, and decision support in a culturally situated manner. Below, we break down how our findings pinpoint broader implications related to HCI-themes of physicality; information curation, visualization, and design; postcoloniality; speculative design; data storytelling; as well as design and methodological implications for the gaps in HCI research that require the domain's prompt attention.}


\subsection{Local Visual Aesthetics and Materiality in Information Curation and Presentation}
\newedits{Our findings show that rural performers relied on existing local conventions of using color, line, shapes, and forms (see section 4 and 5.1)}. For example, to perform for political campaigning events, they put on green and white outfits which are the colors of peace and life in the Bangladeshi context. Additionally, they brought doves as a sign of peace and promises to persuade the villages on behalf of the candidate. These aesthetic and material conventions are culturally situated. They are understandable and relatable to the community members as long as they imagine communionship through shared understanding in common interests, sensibilities of sovereignty, and collective limitations \cite{anderson2006imagined}.


A vast body of literature in information visualization works on case-based appropriate and effective visual attributes, including color, line, shapes, and forms \cite{heer2012color, ward2010interactive, szafir2017modeling, 8017604, heer2012color, du2017isphere, ben2019types}. \newedits{Our work joins this literature and urges that understanding such culturally situated conventions is essential for researchers and visual designers who curate or convey cultural information. This is especially the case for those who work in the space of iconography and design technologies for specific communities and their locally meaningful representational practices;} otherwise, they might risk designing misleading or futile presentations to communities.

\newedits{While it is well known in the data visualization community that, for example, using stop-light colors such as red and green may threaten cross-cultural interpretations (e.g., red may imply wealth rather than undesirability) \cite{tham2020systematic, bazley2021visual}, there is comparably less scholarship on the ways that color, line, shapes, and forms ought to be identified and situated on a local level \cite{kawai2023good, BeijingOperaMask2013, szafir2017modeling, heer2012color, ward2010interactive, ahmad2021red}.} Though there exist general best practices related to factors such as discriminability of color differences \cite{heer2012color}, there is not yet a set of methods for evoking and documenting conventions on a local, situated basis. While there may be no formal canon in the literature, \newedits{our observations demonstrate that there are rich material and oral traditions that may be indexed by motivated designers (see section 4 and 5.1)}. We believe there is a critical need for developing reliable understandings of local context in order to provide meaningful feedback, especially in areas where visual factors may need to be as salient as possible (such as low-literacy regions). Moreover, we demonstrate that ethnographic methods can readily expose such factors which can become general principles for use by data practitioners. 

\subsection{Importance of Physicality in Rural Informational Presentations}
Our work joins threads of ongoing \newedits{physicality research in HCI-design, social computing, and data science}. Physicality has remained a core concern in tangible interface research for almost two decades, where researchers have contributed to application areas such as architecture, engineering, and industrial design by developing compelling mockups for collaborators to share their senses in shared work \cite{brandt2007tangible, hornecker2011role, israel2009intuitive}. Notable social computing works have discussed the use of physicality in collaborative research, where concerns circled around collocation and its benefit through collaborators' position, posture, movement, dynamic references, and sharing artifacts. \cite{olson2002currently, segal1994effects, norman2014things, lerdahl2001staging, tang1991findings, pedersen2020staging}.

A community of visualization, data science, and AI research has engaged with physicalization and related concerns, defining physicality as ``how the shape and materiality properties of artifacts and the behavior of their representation encode data" \cite{hogan2012does, gross2014structures, jansen2015opportunities, offenhuber2020we}. Recent works by Sultana et al. moved forward with these materiality questions regarding visual and data narrative, investigated rural marginalized communities' visual practices, and proposed an alternative visual and data narrative grammar useful to understand and work with such communities that are culturally rich and yet distinctive from modern western notion of data practices \cite{sultana2021seeing, sultana2023communicating}.

Our research advances these areas by bringing insights into how such culturally rich communities organize their information through performances where \newedits{materiality, tradition, spirituality, local myths, and faith-based practices are highly salient (see 5.1, 5.2, and 5.5)}.
We note that elements such as staging and the physiognomy of performers were necessary components in a performance for accomplishing narrative goals and delivering lessons (for example, in identifying antisocial behaviors or character archetypes). As mentioned in the previous section, there is no guarantee that physicality is universal. Rather, it is also situated in local culture and tradition. 
As a practitioner organizes information from visual sources, it is key for them to be able to interpret these subtle physical aspects in their curation process. In turn, practitioners also must be sensitized to the ways in which physical forms also shape interpretation in such culturally rich communities.
For this reason, connecting to our findings is crucial for data scientists and researchers to remain careful about not discarding the cultural sensitivities in data practices, especially as data systems make use of wider varieties of inputs and become increasingly embedded in infrastructure, occupations, and everyday items.



\subsection{Prioritizing Factuality and Emotionality in Performance and Storytelling}
While the majority of today's visualization literature seeks to remain objective and appreciate factuality, we found that our performers, while attentive to factuality in how they organized their information for presentation, also used emotionality in narrative construction and allowed the audiences to have multiple perspectives based on their subjective standing \cite{arnheim1969visual, arnheim1947perceptual, manovich2002data, strothotte2012computational}. For example, \newedits{a participant discussed Bhandari's actual timeline and how long they were active as a performer and educator and informed us about disagreements among the audience (see 5.2)}. They also informed us about the disagreement regarding how long Gazi and Kalu prayed in the forest. However, the participants pointed out that this factual part of this story could be a part of the discussion, but the takeaway should be the moral lessons from these stories that are often teased in the performances through emotionality.

The way rural Bangladeshi information curation allows such factuality, emotionality, and polyvocality in performances and yet is relatable to people with different perspectives could be a lesson for the visual designers working on developing technology for conflicting voices. \newedits{There is a rich intersection between data curation, visual design, and cultural sensitivities at play in our observations.} Current literature at the intersection of emotionality and visualization research focuses mostly on serious storytelling, including healthcare crises, war, and other negative stories \cite{COMICCop96:online, TheProje32:online, 500000do29:online}. Our research extends this literature toward factuality, emotionality, and morality in data curation. We call for researchers to consider ways that both the ground truth used as material for curation and the mechanisms that curation uses to provide new insights might embrace polyvocality and uneasy tensions between factuality and emotionality. Audiences were comfortable working in both modes simultaneously when it matched their situated understanding of the information--- why shouldn't our systems accept these fluid boundaries?


\subsection{Engaging with Missing History}
\newedits{Our study also found that rural Bangladeshi performers often borrowed forms from local myths and folktales (see 5.1, 5.2 and 5.4)}. Their contents and performances are reliant on cultural practices such as stigma and superstitions and religious influences. While previous work in HCI and information visualization have urged HCI, critical data studies, and visualization researchers to engage with local communities through their history, culture, and norms \cite{sultana2021seeing, peck2019data, sayago2016conceptualization}, we also argue that understanding habits related to people's recreational practices is also essential. This is particularly challenging for a community like Bangladesh, which has a long history of being colonized, and as a result, most of its recorded history focuses on political and financial events that were important to colonizers.  However, many communities in Bangladesh have kept their own records of history informally, through folk music, dramas, and games. \newedits{Even so, as these records are no longer practiced, much of Bangladesh's history is missing ~\cite{banerjee2006politics, banerjee2016writing} -- a central post-colonial challenge in HCI. }

This missing history is important to the identity of the communities that created it, and losing it deprives those communities of the benefits of conventions and shared social contexts. If they were readily accessible to both insiders and outsiders, they might help to ease communication and find common ground. This relates to broader questions which are currently in discussion in fields such as indigenous studies \cite{walter2021indigenous} about the stewardship and sovereignty of data sources that are being curated in the potential exclusion of marginalized peoples. We found that rural people in Bangladesh often create visual media that use culturally significant elements, such as mythological and local game components. The significance of these elements is often passed down through generations, and is not a part of recorded history. In this regard, we argue that data curation and visual designers would benefit by directly engaging with the communities through ethnographic and historiographic studies, and design workshops. Visualization researchers frequently co-design with communities \cite{zhang2020,alper2017visualization, sultana2021seeing, peck2019data, bressa2019sketching}. We argue for more such research activism with rural Bangladeshi marginalized populations and other Global South communities. 

\subsection{Implications to Design and Methodologies}
\newedits{While our research offers multifaceted implications for HCI design and methodologies, we discuss here two that we believe require the domain's immediate attention. Our findings show how rural performers draw on histories shaped by incomplete archives, colonial omissions, and deeply affective moral storytelling. We specifically note that the ``sources of data'' in the performances originate from uneven historical and political record-keeping (see 5.4), while audiences evaluate messages through emotional and moral resonance rather than through factual lineage (see 5.5).}

\newedits{This leads to the first design gap: most data-storytelling tools assume linear provenance and singular truth and lack support for plural, layered, and contradictory histories, whereas the communities we studied are comfortable with layered, contradictory, and multiple versions of narratives (e.g., polyvocal accounts). Similar tensions between official data and lived experience appear in participatory environmental sensing projects, where residents used DIY sensing to contest state-produced data and foreground situated interpretation \cite{kuznetsov2011ceci, kuznetsov2011nurturing}. Our findings therefore point toward participatory, speculative, and multimodal design approaches that give communities meaningful control over how their histories and interpretations are represented. One opportunity is the development of community-run data storytelling spaces, inspired by Indigenous data sovereignty infrastructures that emphasize narrative autonomy and collective governance \cite{walter2021indigenous, carroll2019care}. These spaces could host collaborative annotation, storytelling, and interpretation, reflecting the participatory infrastructuring practices through which communities form publics and negotiate shared futures \cite{dantec2013infrastructuring}.}

\newedits{We identify the second design gap by observing how performers and audiences collaboratively interpret, adjust, and morally negotiate stories during live performances, yet existing data systems offer no similar structures for shared narrative governance. Current data practices do not provide mechanisms for community-led governance or interpretive authority. Designers can draw on data humanism, which foregrounds emotion, relational meaning, and subjectivity in visual communication \cite{lupi2017data}, and on work in data physicalization, which uses tangible and sensory forms to support community interpretation and ethical engagement \cite{jansen2015opportunities}. We argue for a move from methods which extract and present information into ones which promote co-creation and shared governance.} 

\newedits{Translating these strategies into rural Bangladeshi contexts, through prototypes such as interactive narrative scrolls, polyvocal timelines, or audio-visual data performances, offers concrete directions for building data-storytelling systems that honor plural truths, foreground embodied sensemaking, and resist extractive representational practices. Hence, we argue that data storytelling and speculative design must actionize and functionalize pluralism, aesthetic resonance, and community-led imagination, not as ancillary values but as core design principles that move HCI beyond fixed narratives and toward systems capable of holding layered, situated, and culturally grounded ways of knowing. In line with Baumer and Silberman's call to recognize when technological interventions risk oversimplifying or even worsening complex social conditions \cite{baumer2011implication}, our findings further suggest that responsible data storytelling in this context may require attending to the situation itself rather than defaulting to a technological ``solution," and in some cases choosing not to design at all.}

\subsection{Limitations and Future Work}
Our work has several limitations which might influence how our implications and broader considerations ought to be interpreted. First, our work is not free from participation bias and selection bias. We engaged with the villagers through a snowball sampling process. While we can assume that opinions and arguments represent the collective view of the residents of the whole geographic area, this may not actually be the case. Second, our interaction with the participants could also have suffered from experimental and methodological flaws. Question-order bias, self-presentation maintenance, and power imbalances all might influence what we observed.

\newedits{This study opens several avenues for future research that build directly on our contributions: } \newedits{First, extending our empirical account of rural Bangladeshi performance traditions and their narrative strategies, future work will co-design and prototype interactive narrative scrolls, polyvocal timelines, and audio-visual data performances with local artists. These prototypes will operationalize our identified narrative structures, aesthetic cues, and interpretive rhythms identified and explore how they translate into computational forms grounded in cultural context.}

\newedits{Second, building on local visual and material storytelling practices and on the interplay of fact, faith, and figure, we will develop a design framework for plural narrative infrastructures. This framework will address the two gaps we identified: limited support for layered and contradictory histories and the absence of mechanisms for shared narrative governance. It will outline interaction patterns that enable collective interpretation and culturally grounded narrative blending and will explore how multimodal cues and embodied techniques from rural performances can be functionalized within interactive systems.}

\newedits{Third, another direction involves developing community-centered data infrastructures, such as local repositories, community-governed annotation hubs, and collaborative archiving practices. These infrastructures foreground local expertise and advance decolonial approaches to AI by prioritizing narrative sovereignty, shared stewardship, and cultural control of data. Since our findings show that performers and audiences already sustain sophisticated systems of narrative preservation and moral interpretation, such infrastructures can support and strengthen practices already active within the community.}

\newedits{Fourth, we will refine our methodological contributions through iterative participatory workshops that integrate fieldnotes, translation challenges, and oral histories into design processes and examine how digital platforms shape emerging performance practices. Finally, we will explore circumstances in which minimal or non-design interventions are more appropriate, ensuring that future technological engagements remain aligned with the situated cultural, ethical, and epistemic foundations documented in this study.}

\section{Conclusion}
This paper shares the results of a ten-month ethnographic study examining how communities utilize entertainment and cultural practices such as \textit{Puthi}, \textit{Bhandari Gaan}, and \textit{Pot music} to curate information. These practices are employed to communicate traditional moral lessons and history. Through our fieldwork, we discovered how these communities embrace polyvocality and incorporate multiple ethical frameworks into their performances. They skillfully construct narratives that blend factuality, emotionality, and aesthetics, adapting their presentations to accommodate evolving technology and audience preferences. The findings of our research are significant for future data-driven and visualization systems to build more inclusive and sustainable data-driven ecosystem by addressing the methods and needs of marginalized communities. 

\begin{acks}
We thank the participants for generously sharing their time, experiences, and insights throughout the study. We also thank the Rural Reconstruction Foundation (RRF) for their crucial assistance with community access and participant recruitment. This work was supported by Sharifa Sultana's ICR fund \#C1-F200250-O434053-P434533.
\end{acks}




\bibliographystyle{ACM-Reference-Format}
\bibliography{0.main}

@article{kawai2023good,
  title={The good, the bad, and the red: implicit color-valence associations across cultures},
  author={Kawai, Claudia and Zhang, Yang and Luk{\'a}cs, G{\'a}sp{\'a}r and Chu, Wenyi and Zheng, Chaoyi and Gao, Cijun and Gozli, Davood and Wang, Yonghui and Ansorge, Ulrich},
  journal={Psychological Research},
  volume={87},
  number={3},
  pages={704--724},
  year={2023},
  publisher={Springer}
}

@misc{BeijingOperaMask2013,
  title        = {History of the Peking Opera Mask},
  author       = {{Beijing Municipal Bureau of Culture and Tourism}},
  year         = {2013},
  month        = {Aug},
  day          = {13},
  howpublished = {\url{https://english.visitbeijing.com.cn/article/47OMwaQB3hD}},
  note         = {Accessed: 2025-12-02}
}

@article{szafir2017modeling,
  title={Modeling color difference for visualization design},
  author={Szafir, Danielle Albers},
  journal={IEEE transactions on visualization and computer graphics},
  volume={24},
  number={1},
  pages={392--401},
  year={2017},
  publisher={IEEE}
}

@inproceedings{heer2012color,
  title={Color naming models for color selection, image editing and palette design},
  author={Heer, Jeffrey and Stone, Maureen},
  booktitle={Proceedings of the SIGCHI Conference on Human Factors in Computing Systems},
  pages={1007--1016},
  year={2012}
}

@inproceedings{ahmad2021red,
  title={When red means good, bad, or Canada: exploring people’s reasoning for choosing color palettes},
  author={Ahmad, Jarryullah and Huynh, Elaine and Chevalier, Fanny},
  booktitle={2021 IEEE Visualization Conference (VIS)},
  pages={56--60},
  year={2021},
  organization={IEEE}
}

@article{carroll2019care,
  title={Die CARE-Prinzipien f{\"u}r indigene Data Governance},
  author={Carroll, Stephanie R and Hudson, Maui and Chapman, Jan and Figueroa-Rodr{\i}guez, OL and Holbrook, Jarita and Lovett, Ray and Materechera, Simeon and Parsons, Mark and Raseroka, Kay and Rodriguez-Lonebear, Desi and others},
  journal={Research Data Alliance International Indigenous Data Sovereignty Interest Group},
  year={2019}
}

@article{walter2021indigenous,
  title={Indigenous data sovereignty in the era of big data and open data},
  author={Walter, Maggie and Lovett, Raymond and Maher, Bobby and Williamson, Bhiamie and Prehn, Jacob and Bodkin-Andrews, Gawaian and Lee, Vanessa},
  journal={Australian Journal of Social Issues},
  volume={56},
  number={2},
  pages={143--156},
  year={2021},
  publisher={Wiley Online Library}
}

@article{dantec2013infrastructuring,
  title={Infrastructuring and the formation of publics in participatory design},
  author={Dantec, Christopher A Le and DiSalvo, Carl},
  journal={Social Studies of Science},
  volume={43},
  number={2},
  pages={241--264},
  year={2013},
  publisher={SAGE Publications Sage UK: London, England}
}

@inproceedings{kuznetsov2011ceci,
  title={Ceci n'est pas une pipe bombe: authoring urban landscapes with air quality sensors},
  author={Kuznetsov, Stacey and Davis, George and Cheung, Jian and Paulos, Eric},
  booktitle={Proceedings of the sigchi conference on human factors in computing systems},
  pages={2375--2384},
  year={2011}
}

@inproceedings{kuznetsov2011nurturing,
  title={Nurturing natural sensors},
  author={Kuznetsov, Stacey and Odom, William and Pierce, James and Paulos, Eric},
  booktitle={Proceedings of the 13th international conference on Ubiquitous computing},
  pages={227--236},
  year={2011}
}

@article{lupi2017data,
  title={Data humanism: the revolutionary future of data visualization},
  author={Lupi, Giorgia},
  journal={Print Magazine},
  volume={30},
  number={3},
  pages={2},
  year={2017}
}

@inproceedings{baumer2011implication,
  title={When the implication is not to design (technology)},
  author={Baumer, Eric PS and Silberman, M Six},
  booktitle={Proceedings of the SIGCHI Conference on Human Factors in Computing Systems},
  pages={2271--2274},
  year={2011}
}

@article{mondal2022discourse,
  title={A Discourse on Patachitra Art with narratives and songs in religious and cultural Scenario of West Bengal.},
  author={Mondal, Shyamal},
  journal={International Journal of Advances in Engineering and Management},
  volume={4},
  number={2},
  pages={183--188},
  year={2022}
}

@techreport{johnston1997real,
  title={Real-time digital libraries based on widely distributed, high performance management of large-data-objects},
  author={Johnston, William and Larsen, C and Lee, Jason and Hoo, Gary and Guojun, J and Tierney, BL and Thompson, M and Terdiman, Joseph},
  year={1997},
  institution={SCAN-9902019}
}

@inproceedings{gray2002online,
  title={Online scientific data curation, publication, and archiving},
  author={Gray, Jim and Szalay, Alexander S and Thakar, Ani R and Stoughton, Christopher and others},
  booktitle={Virtual observatories},
  volume={4846},
  pages={103--107},
  year={2002},
  organization={SPIE}
}

@article{shreeves2008introduction,
  title={Introduction: Institutional repositories: Current state and future},
  author={Shreeves, Sarah L and Cragin, Melissa H},
  journal={Library Trends},
  volume={57},
  number={2},
  pages={89--97},
  year={2008},
  publisher={Johns Hopkins University Press}
}

@article{palmer2013foundations,
  title={Foundations of data curation: The pedagogy and practice of “purposeful work” with research data},
  author={Palmer, Carole L and Weber, Nicholas M and Mu{\~n}oz, Trevor and Renear, Allen H},
  journal={Archive Journal},
  volume={3},
  year={2013}
}

@article{johnston2018important,
  title={How important is data curation? Gaps and opportunities for academic libraries},
  author={Johnston, Lisa R and Carlson, Jacob and Hudson-Vitale, Cynthia and Imker, Heidi and Kozlowski, Wendy and Olendorf, Robert and Stewart, Claire},
  journal={Journal of Librarianship and Scholarly Communication},
  volume={6},
  number={1},
  year={2018},
  publisher={Iowa State University Digital Press}
}

@article{weber2012current,
  title={Current trends and future directions in data curation research and education},
  author={Weber, Nicholas M and Palmer, Carole L and Chao, Tiffany C},
  journal={Journal of Web Librarianship},
  volume={6},
  number={4},
  pages={305--320},
  year={2012},
  publisher={Taylor \& Francis}
}

@article{cragin2007educational,
  title={An educational program on data curation},
  author={Cragin, Melissa H and Heidorn, P Bryan and Palmer, Carole L and Smith, Linda C},
  year={2007}
}

@article{gold2010data,
  title={Data curation and libraries: Short-term developments, long-term prospects},
  author={Gold, Anna},
  year={2010}
}

@misc{DataCura83:online,
author = {Data Science at NIH},
title = {Data Curation Network – Event Series (ODSS, NLM)},
howpublished = {\url{https://datascience.nih.gov/data-curation-network-event-series}},
year = {2023}
}

@misc{CENDIDat6:online,
author = {CENDI},
title = {Data Curation Working Group},
howpublished = {\url{https://www.cendi.gov/working-groups/data-curation.shtml}},
year = {2023}
}

@article{antognoli2020reproducibility,
  title={Reproducibility literature analysis-a federal information professional perspective},
  author={Antognoli, Erin and Avila, Regina L and Sears, Jonathan and Christiansen, Leighton L and Tieman, Jessica and Hart, Jacquelyn},
  journal={IASSIST Quarterly},
  volume={44},
  number={1-2},
  pages={1--26},
  year={2020}
}

@misc{BTSDataS30:online,
author = {US Department of Transportation},
title = {BTS Data Services and Data Curation},
howpublished = {\url{https://www.transportation.gov/government/traffic-records/bts-data-services-and-data-curation}},
year = {2023}
}

@misc{Aboutthe22:online,
author = {US EPA},
title = {About the Scientific Computing and Data Curation Division},
howpublished = {\url{https://www.epa.gov/aboutepa/about-scientific-computing-and-data-curation-division}},
year = {2023}
}

@misc{DataCata85:online,
author = {Institute of Museum and Library Services},
title = {Data Catalog},
howpublished = {\url{https://www.imls.gov/research-tools/data-collection}},
year = {2023}
}

@misc{GooglePu52:online,
author = {Google},
title = {Google Public Data Explorer},
howpublished = {\url{https://www.google.com/publicdata/directory}},
year = {2023}
}

@misc{PagePubl73:online,
author = {Graph API},
title = {Page Public Content Access},
howpublished = {\url{https://developers.facebook.com/docs/features-reference/page-public-content-access/}},
year = {Facebook}
}

@article{folbre1989women,
  title={Women's work and women's households: Gender bias in the US census},
  author={Folbre, Nancy and Abel, Marjorie},
  journal={Social Research},
  pages={545--569},
  year={1989},
  publisher={JSTOR}
}

@article{wachter2000fifth,
  title={The fifth cell: correlation bias in US census adjustment},
  author={Wachter, Kenneth W and Freedman, David A},
  journal={Evaluation Review},
  volume={24},
  number={2},
  pages={191--211},
  year={2000},
  publisher={Sage Publications Sage CA: Thousand Oaks, CA}
}

@article{williams2015racial,
  title={Racial bias in health care and health: challenges and opportunities},
  author={Williams, David R and Wyatt, Ronald},
  journal={Jama},
  volume={314},
  number={6},
  pages={555--556},
  year={2015},
  publisher={American Medical Association}
}

@article{grundmeier2015imputing,
  title={Imputing missing race/ethnicity in pediatric electronic health records: reducing bias with use of US census location and surname data},
  author={Grundmeier, Robert W and Song, Lihai and Ramos, Mark J and Fiks, Alexander G and Elliott, Marc N and Fremont, Allen and Pace, Wilson and Wasserman, Richard C and Localio, Russell},
  journal={Health services research},
  volume={50},
  number={4},
  pages={946--960},
  year={2015},
  publisher={Wiley Online Library}
}

@article{donovan2019toward,
  title={Toward a more humane genetics education: Learning about the social and quantitative complexities of human genetic variation research could reduce racial bias in adolescent and adult populations},
  author={Donovan, Brian M and Semmens, Rob and Keck, Phillip and Brimhall, Elizabeth and Busch, KC and Weindling, Monica and Duncan, Alex and Stuhlsatz, Molly and Bracey, Zo{\"e} Buck and Bloom, Mark and others},
  journal={Science Education},
  volume={103},
  number={3},
  pages={529--560},
  year={2019},
  publisher={Wiley Online Library}
}

@book{burton1994burdens,
  title={Burdens of history: British feminists, Indian women, and imperial culture, 1865-1915},
  author={Burton, Antoinette M},
  year={1994},
  publisher={Univ of North Carolina Press}
}

@book{dirks2006scandal,
  title={The scandal of empire: India and the creation of imperial Britain},
  author={Dirks, Nicholas B},
  year={2006},
  publisher={Harvard University Press}
}

@article{yuan2022occupational,
  title={Occupational stress and health risk of employees working in the garments sector of Bangladesh: An empirical study},
  author={Yuan, Deli and Gazi, Md Abu Issa and Rahman, Md Alinoor and Dhar, Bablu Kumar and Rahaman, Md Atikur},
  journal={Frontiers in public health},
  volume={10},
  pages={938248},
  year={2022},
  publisher={Frontiers}
}

@article{prentice2018health,
  title={Health and safety in garment workers’ lives: Setting a new research agenda},
  author={Prentice, Rebecca and De Neve, Geert and Mezzadri, Alessandra and Ruwanpura, Kanchana N},
  journal={Geoforum},
  volume={88},
  pages={157--160},
  year={2018},
  publisher={Elsevier}
}

@inproceedings{dias2020data,
  title={Data curation: towards a tool for all},
  author={Dias, Jos{\'e} and Cunha, J{\'a}come and Pereira, Rui},
  booktitle={International Conference on Human-Computer Interaction},
  pages={176--183},
  year={2020},
  organization={Springer}
}

@inproceedings{rezig2019towards,
  title={Towards an end-to-end human-centric data cleaning framework},
  author={Rezig, El Kindi and Ouzzani, Mourad and Elmagarmid, Ahmed K and Aref, Walid G and Stonebraker, Michael},
  booktitle={Proceedings of the Workshop on Human-In-the-Loop Data Analytics},
  pages={1--7},
  year={2019}
}

@inproceedings{krishnan2016towards,
  title={Towards reliable interactive data cleaning: A user survey and recommendations},
  author={Krishnan, Sanjay and Haas, Daniel and Franklin, Michael J and Wu, Eugene},
  booktitle={Proceedings of the Workshop on Human-In-the-Loop Data Analytics},
  pages={1--5},
  year={2016}
}

@inproceedings{tong2014crowdcleaner,
  title={Crowdcleaner: Data cleaning for multi-version data on the web via crowdsourcing},
  author={Tong, Yongxin and Cao, Caleb Chen and Zhang, Chen Jason and Li, Yatao and Chen, Lei},
  booktitle={2014 IEEE 30th International Conference on Data Engineering},
  pages={1182--1185},
  year={2014},
  organization={IEEE}
}

@article{greenwald2022whole,
  title={Whole-cell segmentation of tissue images with human-level performance using large-scale data annotation and deep learning},
  author={Greenwald, Noah F and Miller, Geneva and Moen, Erick and Kong, Alex and Kagel, Adam and Dougherty, Thomas and Fullaway, Christine Camacho and McIntosh, Brianna J and Leow, Ke Xuan and Schwartz, Morgan Sarah and others},
  journal={Nature biotechnology},
  volume={40},
  number={4},
  pages={555--565},
  year={2022},
  publisher={Nature Publishing Group US New York}
}

@book{monarch2021human,
  title={Human-in-the-Loop Machine Learning: Active learning and annotation for human-centered AI},
  author={Monarch, Robert Munro},
  year={2021},
  publisher={Simon and Schuster}
}

@article{van2021biological,
  title={Biological data annotation via a human-augmenting AI-based labeling system},
  author={van der Wal, Douwe and Jhun, Iny and Laklouk, Israa and Nirschl, Jeff and Richer, Lara and Rojansky, Rebecca and Theparee, Talent and Wheeler, Joshua and Sander, J{\"o}rg and Feng, Felix and others},
  journal={NPJ digital medicine},
  volume={4},
  number={1},
  pages={145},
  year={2021},
  publisher={Nature Publishing Group UK London}
}

@inproceedings{chen2020building,
  title={Building data curation processes with crowd intelligence},
  author={Chen, Tianwa and Han, Lei and Demartini, Gianluca and Indulska, Marta and Sadiq, Shazia},
  booktitle={Advanced Information Systems Engineering: CAiSE Forum 2020, Grenoble, France, June 8--12, 2020, Proceedings 32},
  pages={29--42},
  year={2020},
  organization={Springer}
}

@inproceedings{vitale2020data,
  title={Data dashboard: exploring centralization and customization in personal data curation},
  author={Vitale, Francesco and Chen, Janet and Odom, William and McGrenere, Joanna},
  booktitle={Proceedings of the 2020 ACM Designing Interactive Systems Conference},
  pages={311--326},
  year={2020}
}

@inproceedings{abdul2018trends,
  title={Trends and trajectories for explainable, accountable and intelligible systems: An hci research agenda},
  author={Abdul, Ashraf and Vermeulen, Jo and Wang, Danding and Lim, Brian Y and Kankanhalli, Mohan},
  booktitle={Proceedings of the 2018 CHI conference on human factors in computing systems},
  pages={1--18},
  year={2018}
}

@article{lage2011receptivity,
  title={Receptivity to library involvement in scientific data curation: A case study at the University of Colorado Boulder},
  author={Lage, Kathryn and Losoff, Barbara and Maness, Jack},
  journal={portal: Libraries and the Academy},
  volume={11},
  number={4},
  pages={915--937},
  year={2011},
  publisher={Johns Hopkins University Press}
}

@article{freitas2016big,
  title={Big data curation},
  author={Freitas, Andr{\'e} and Curry, Edward},
  journal={New horizons for a data-driven economy: A roadmap for usage and exploitation of big data in Europe},
  pages={87--118},
  year={2016},
  publisher={Springer International Publishing}
}

@article{pouchard2015revisiting,
  title={Revisiting the data lifecycle with big data curation},
  author={Pouchard, Line},
  year={2015}
}

@article{rycroft2006santalism,
  title={Santalism: reconfiguring ‘the Santal’ in Indian art and politics},
  author={Rycroft, Daniel J},
  journal={Indian Historical Review},
  volume={33},
  number={1},
  pages={150--174},
  year={2006},
  publisher={Sage Publications Sage India: New Delhi, India}
}

@article{toffin2009janajati,
  title={The janajati/adivasi movement in Nepal: Myths and realities of indigeneity},
  author={Toffin, G{\'e}rard},
  journal={Sociological Bulletin},
  volume={58},
  number={1},
  pages={25--42},
  year={2009},
  publisher={SAGE Publications Sage India: New Delhi, India}
}

@article{prasad2006unravelling,
  title={Unravelling the Forms of ‘Adivasi’ Organization and Resistance in Colonial India},
  author={Prasad, Archana},
  journal={Indian Historical Review},
  volume={33},
  number={1},
  pages={225--244},
  year={2006},
  publisher={SAGE Publications Sage India: New Delhi, India}
}

@article{baker2009data,
  title={Data stewardship: Environmental data curation and a web-of-repositories},
  author={Baker, Karen S and Yarmey, Lynn},
  year={2009}
}

@misc{SignsonT9:online,
author = {Merlini, Marco},
title = {Signs on Tartaria Tablets found in the Romanian folkloric art},
howpublished = {\url{http://www.prehistory.it/ftp/arta_populara01.htm}},
year = {2023}
}

@article{huylebrouck2019missing,
  title={Missing link},
  author={Huylebrouck, Dirk and Huylebrouck, Dirk},
  journal={Africa and Mathematics: From Colonial Findings Back to the Ishango Rods},
  pages={153--166},
  year={2019},
  publisher={Springer}
}

@article{haigh2016charles,
  title={How Charles Bachman invented the DBMS, a foundation of our digital world},
  author={Haigh, Thomas},
  journal={Communications of the ACM},
  volume={59},
  number={7},
  pages={25--30},
  year={2016},
  publisher={ACM New York, NY, USA}
}

@article{bachman2009origin,
  title={The origin of the integrated data store (IDS): The first direct-access DBMS},
  author={Bachman, Charles W},
  journal={IEEE Annals of the History of Computing},
  volume={31},
  number={4},
  pages={42--54},
  year={2009},
  publisher={IEEE}
}

@inproceedings{bachman1975trends,
  title={Trends in database management: 1975},
  author={Bachman, Charles W},
  booktitle={Proceedings of the May 19-22, 1975, national computer conference and exposition},
  pages={569--576},
  year={1975}
}

@article{stierhoff1998history,
  title={A history of the IBM Systems Journal},
  author={Stierhoff, George C and Davis, Alfred G},
  journal={IEEE Annals of the History of Computing},
  volume={20},
  number={1},
  pages={29--35},
  year={1998},
  publisher={IEEE}
}

@article{Corbi1989IBM,
  title={Program Understanding: Challenge for the 1990s},
  author={Corbi, Thomas A.},
  journal={IBM Systems Journal},
  volume={28},
  number={2},
  pages={294},
  year={1989},
  publisher={IBM},
}

@inproceedings{lord2004data,
  title={From data deluge to data curation},
  author={Lord, Philip and Macdonald, Alison and Lyon, Liz and Giaretta, David},
  booktitle={Proceedings of the UK e-science All Hands meeting},
  volume={440},
  number={2004},
  pages={371--375},
  year={2004},
  organization={Citeseer}
}

@article{curry2010role,
  title={The role of community-driven data curation for enterprises},
  author={Curry, Edward and Freitas, Andre and O’Ri{\'a}in, Sean},
  journal={Linking enterprise data},
  pages={25--47},
  year={2010},
  publisher={Springer}
}

@article{karasti2006enriching,
  title={Enriching the notion of data curation in e-science: data managing and information infrastructuring in the long term ecological research (LTER) network},
  author={Karasti, Helena and Baker, Karen S and Halkola, Eija},
  journal={Computer Supported Cooperative Work (CSCW)},
  volume={15},
  pages={321--358},
  year={2006},
  publisher={Springer}
}

@article{witt2009constructing,
  title={Constructing data curation profiles},
  author={Witt, Michael and Carlson, Jacob and Brandt, D Scott and Cragin, Melissa H},
  journal={International Journal of Digital Curation},
  volume={4},
  number={3},
  pages={93--103},
  year={2009}
}

@article{victorelli2020understanding,
  title={Understanding human-data interaction: Literature review and recommendations for design},
  author={Victorelli, Eliane Zambon and Dos Reis, Julio Cesar and Hornung, Heiko and Prado, Alysson Bolognesi},
  journal={International journal of human-computer studies},
  volume={134},
  pages={13--32},
  year={2020},
  publisher={Elsevier}
}

@article{roh2019survey,
  title={A survey on data collection for machine learning: a big data-ai integration perspective},
  author={Roh, Yuji and Heo, Geon and Whang, Steven Euijong},
  journal={IEEE Transactions on Knowledge and Data Engineering},
  volume={33},
  number={4},
  pages={1328--1347},
  year={2019},
  publisher={IEEE}
}

@article{chen2022merging,
  title={Merging data curation and machine learning to improve nanomedicines},
  author={Chen, Chen and Yaari, Zvi and Apfelbaum, Elana and Grodzinski, Piotr and Shamay, Yosi and Heller, Daniel A},
  journal={Advanced drug delivery reviews},
  volume={183},
  pages={114172},
  year={2022},
  publisher={Elsevier}
}

@article{ethics,
  title =        {{Social Impacts of Computing: Codes of Professional
                  Ethics}},
  author =       {R. E. Anderson},
  doi =          "10.1177/089443939201000402",
  journal =      "Social Science Computer Review December",
  year =         1992,
  volume =       10,
  number =       4,
  pages =        "453-469"
}

@article{olson2002currently,
  title={The (currently) unique advantages of collocated work},
  author={Olson, Judith S and Teasley, Stephanie and Covi, Lisa and Olson, Gary},
  journal={Distributed work},
  pages={113--135},
  year={2002},
  publisher={MIT Press, Cambridge, MA}
}

@techreport{segal1994effects,
  title={Effects of checklist interface on non-verbal crew communications},
  author={Segal, Leon D},
  year={1994}
}

@book{norman2014things,
  title={Things that make us smart: Defending human attributes in the age of the machine},
  author={Norman, Don},
  year={2014},
  publisher={Diversion Books}
}

@book{lerdahl2001staging,
  title={Staging for creative collaboration in design teams},
  author={Lerdahl, Erik},
  year={2001},
  publisher={Fakultet for ingeni{\o}rvitenskap og teknologi}
}

@article{pedersen2020staging,
  title={Staging negotiation spaces: A co-design framework},
  author={Pedersen, Signe},
  journal={Design Studies},
  volume={68},
  pages={58--81},
  year={2020},
  publisher={Elsevier}
}

@article{tang1991findings,
  title={Findings from observational studies of collaborative work},
  author={Tang, John C},
  journal={International Journal of Man-machine studies},
  volume={34},
  number={2},
  pages={143--160},
  year={1991},
  publisher={Elsevier}
}

@article{brandt2007tangible,
  title={How tangible mock-ups support design collaboration},
  author={Brandt, Eva},
  journal={Knowledge, Technology \& Policy},
  volume={20},
  number={3},
  pages={179--192},
  year={2007},
  publisher={Springer}
}

@article{hornecker2011role,
  title={The role of physicality in tangible and embodied interactions},
  author={Hornecker, Eva},
  journal={interactions},
  volume={18},
  number={2},
  pages={19--23},
  year={2011},
  publisher={ACM New York, NY, USA}
}

@inproceedings{hogan2012does,
  title={How does representation modality affect user-experience of data artifacts?},
  author={Hogan, Trevor and Hornecker, Eva},
  booktitle={Haptic and Audio Interaction Design: 7th International Conference, HAID 2012, Lund, Sweden, August 23-24, 2012. Proceedings 7},
  pages={141--151},
  year={2012},
  organization={Springer}
}

@article{gross2014structures,
  title={Structures, forms, and stuff: the materiality and medium of interaction},
  author={Gross, Shad and Bardzell, Jeffrey and Bardzell, Shaowen},
  journal={Personal and Ubiquitous Computing},
  volume={18},
  pages={637--649},
  year={2014},
  publisher={Springer}
}

@inproceedings{jansen2015opportunities,
  title={Opportunities and challenges for data physicalization},
  author={Jansen, Yvonne and Dragicevic, Pierre and Isenberg, Petra and Alexander, Jason and Karnik, Abhijit and Kildal, Johan and Subramanian, Sriram and Hornb{\ae}k, Kasper},
  booktitle={Proceedings of the 33rd Annual ACM Conference on Human Factors in Computing Systems},
  pages={3227--3236},
  year={2015}
}

@article{offenhuber2020we,
  title={What we talk about when we talk about data physicality},
  author={Offenhuber, Dietmar},
  journal={IEEE Computer Graphics and Applications},
  volume={40},
  number={6},
  pages={25--37},
  year={2020},
  publisher={IEEE}
}

@article{israel2009intuitive,
  title={On intuitive use, physicality and tangible user interfaces},
  author={Israel, Johann H and Hurtienne, Jorn and Pohlmeyer, Anna E and Mohs, Carsten and Kindsmuller, Martin and Naumann, Anja},
  journal={International Journal of Arts and Technology},
  volume={2},
  number={4},
  pages={348--366},
  year={2009},
  publisher={Inderscience Publishers}
}

@article{sultana2021seeing,
  title={Seeing in context: Traditional visual communication practices in rural bangladesh},
  author={Sultana, Sharifa and Ahmed, Syed Ishtiaque and Rzeszotarski, Jeffrey M},
  journal={Proceedings of the ACM on Human-Computer Interaction},
  volume={4},
  number={CSCW3},
  pages={1--31},
  year={2021},
  publisher={ACM New York, NY, USA}
}

@inproceedings{sultana2023communicating,
  title={Communicating Consequences: Visual Narratives, Abstraction, and Polysemy in Rural Bangladesh},
  author={Sultana, Sharifa and Ahmed, Syed Ishtiaque and Rzeszotarski, Jeffrey M},
  booktitle={Proceedings of the 2023 CHI Conference on Human Factors in Computing Systems},
  pages={1--19},
  year={2023}
}

@article{segel2010narrative,
  title={Narrative visualization: Telling stories with data},
  author={Segel, Edward and Heer, Jeffrey},
  journal={IEEE transactions on visualization and computer graphics},
  volume={16},
  number={6},
  pages={1139--1148},
  year={2010},
  publisher={IEEE}
}

@inproceedings{bressa2019sketching,
  title={Sketching and ideation activities for situated visualization design},
  author={Bressa, Nathalie and Wannamaker, Kendra and Korsgaard, Henrik and Willett, Wesley and Vermeulen, Jo},
  booktitle={Proceedings of the 2019 on Designing Interactive Systems Conference},
  pages={173--185},
  year={2019}
}

@book{anderson2006imagined,
  title={Imagined communities: Reflections on the origin and spread of nationalism},
  author={Anderson, Benedict},
  year={2006},
  publisher={Verso books}
}

@article{arnheim1969visual,
  title={Visual Thinking University of California Press},
  author={Arnheim, Rudolf},
  journal={Berkeley and Los Angeles},
  year={1969}
}

@article{arnheim1947perceptual,
  title={Perceptual abstraction and art.},
  author={Arnheim, Rudolf},
  journal={Psychological Review},
  volume={54},
  number={2},
  pages={66},
  year={1947},
  publisher={American Psychological Association}
}

@article{manovich2002data,
  title={Data visualization as new abstraction and anti-sublime},
  author={Manovich, Lev},
  journal={Small tech: The culture of digital tools},
  volume={3},
  number={9},
  year={2002},
  publisher={U of Minnesota Press}
}

@book{strothotte2012computational,
  title={Computational Visualization: Graphics, abstraction and interactivity},
  author={Strothotte, Thomas},
  year={2012},
  publisher={Springer Science \& Business Media}
}

@book{huff1993lie,
  title={How to lie with statistics},
  author={Huff, Darrell},
  year={1993},
  publisher={WW Norton \& Company}
}

@misc{Visualiz1:online,
author = {Valle, Mario},
title = {Visualization and art},
howpublished = {\url{http://mariovalle.name/visualization/VizArt.html}},
month = {March},
year = {2005}
}

@inproceedings{van2005value,
  title={The value of visualization},
  author={Van Wijk, Jarke J},
  booktitle={VIS 05. IEEE Visualization, 2005.},
  pages={79--86},
  year={2005},
  organization={IEEE}
}

@article{carpendale2017subjectivity,
  title={Subjectivity in personal storytelling with visualization},
  author={Carpendale, Sheelagh and Thudt, Alice and Perin, Charles and Willett, Wesley},
  journal={Information Design Journal},
  volume={23},
  number={1},
  pages={48--64},
  year={2017},
  publisher={John Benjamins}
}

@article{meyer2019criteria,
  title={Criteria for rigor in visualization design study},
  author={Meyer, Miriah and Dykes, Jason},
  journal={IEEE transactions on visualization and computer graphics},
  volume={26},
  number={1},
  pages={87--97},
  year={2019},
  publisher={IEEE}
}

@article{tufte1985visual,
  title={The visual display of quantitative information},
  author={Tufte, Edward R},
  journal={The Journal for Healthcare Quality (JHQ)},
  volume={7},
  number={3},
  pages={15},
  year={1985},
  publisher={LWW}
}

@article{jgrgenson1995visualization,
  title={Is Visualization Struggling under the Myth of Objectivity?},
  author={JGrgenson, Loki and Kriz, BC Ronald and Tech, Virgina and Mones-Hattal, Barbara},
  year={1995}
}

@incollection{ambrosio2014objectivity,
  title={Objectivity and representative practices across artistic and scientific visualization},
  author={Ambrosio, Chiara},
  booktitle={Visualization in the age of computerization},
  pages={118--144},
  year={2014},
  publisher={Routledge}
}

@MISC{34RRF,
   author = {{Rural Reconstruction Foundation}},
   howpublished = {\url{http://www.rrf-bd.org/}},
  year={2017}
}

@article{75strauss1990open,
  title={Open coding},
  author={Strauss, Anselm and Corbin, Juliet},
  journal={Basics of qualitative research: Grounded theory procedures and techniques},
  volume={2},
  number={1990},
  pages={101--121},
  year={1990}
}

@book{76boyatzis1998transforming,
  title={Transforming qualitative information: Thematic analysis and code development},
  author={Boyatzis, Richard E},
  year={1998},
  publisher={sage}
}

@article{banerjee2016writing,
  title={Writing the Adivasi: Some historiographical notes},
  author={Banerjee, Prathama},
  journal={The Indian Economic \& Social History Review},
  volume={53},
  number={1},
  pages={131--153},
  year={2016},
  publisher={SAGE Publications Sage India: New Delhi, India}
}

@article{banerjee2006politics,
  title={Politics of Time:'primitives' and History-writing in a Colonial Society},
  author={Banerjee, Prathama},
  year={2006}
}

@book{ward2010interactive,
  title={Interactive data visualization: foundations, techniques, and applications},
  author={Ward, Matthew O and Grinstein, Georges and Keim, Daniel},
  year={2010},
  publisher={CRC Press}
}

@inproceedings{8017604,  
author={Szafir, Danielle A.},  
booktitle={IEEE Transactions on Visualization and Computer Graphics},   
title={Modeling Color Difference for Visualization Design},   year={2018},  
volume={24},  
number={1},  
pages={392-401}
}

@inproceedings{alper2017visualization,
  title={Visualization literacy at elementary school},
  author={Alper, Basak and Riche, Nathalie Henry and Chevalier, Fanny and Boy, Jeremy and Sezgin, Metin},
  booktitle={Proceedings of the 2017 CHI Conference on Human Factors in Computing Systems},
  pages={5485--5497},
  year={2017}
}

@inproceedings{zhang2020,
author = {Zhang, Jiayi Eris and Sultanum, Nicole and Bezerianos, Anastasia and Chevalier, Fanny},
title = {DataQuilt: Extracting Visual Elements from Images to Craft Pictorial Visualizations},
year = {2020},
isbn = {9781450367080},
publisher = {Association for Computing Machinery},
address = {New York, NY, USA},
url = {https://doi.org/10.1145/3313831.3376172},
doi = {10.1145/3313831.3376172},
booktitle = {Proceedings of the 2020 CHI Conference on Human Factors in Computing Systems},
pages = {1–13},
numpages = {13},
location = {Honolulu, HI, USA},
series = {CHI ’20}
}

@inproceedings{du2017isphere,
  title={isphere: Focus+ context sphere visualization for interactive large graph exploration},
  author={Du, Fan and Cao, Nan and Lin, Yu-Ru and Xu, Panpan and Tong, Hanghang},
  booktitle={Proceedings of the 2017 CHI Conference on Human Factors in Computing Systems},
  pages={2916--2927},
  year={2017}
}

@inproceedings{ben2019types,
  title={Types of graphic interface design and their role in learning via mathematical applets at the elementary school},
  author={Ben-Haim, Eitan and Cohen, Anat and Tabach, Michal},
  booktitle={Eleventh Congress of the European Society for Research in Mathematics Education},
  number={3},
  year={2019},
  organization={Freudenthal Group; Freudenthal Institute; ERME}
}

@inproceedings{peck2019data,
  title={Data is Personal: Attitudes and Perceptions of Data Visualization in Rural Pennsylvania},
  author={Peck, Evan M and Ayuso, Sofia E and El-Etr, Omar},
  booktitle={Proceedings of the 2019 CHI Conference on Human Factors in Computing Systems},
  pages={1--12},
  year={2019}
}

@article{Biernacki1981,
author = {Biernacki, P. and Waldorf, D.},
journal = {Sociological Methods {\&} Research},
pages = {141--163},
title = {{Snowball Sampling: Problems and Techniques of Chain Referral Sampling}},
volume = {10},
year = {1981}
}

@article{wang2019emotional,
  title={An emotional response to the value of visualization},
  author={Wang, Yun and Segal, Adrien and Klatzky, Roberta and Keefe, Daniel F and Isenberg, Petra and Hurtienne, J{\"o}rn and Hornecker, Eva and Dwyer, Tim and Barrass, Stephen},
  journal={IEEE computer graphics and applications},
  volume={39},
  number={5},
  pages={8--17},
  year={2019},
  publisher={IEEE}
}

@inproceedings{lan2022negative,
  title={Negative emotions, positive outcomes? exploring the communication of negativity in serious data stories},
  author={Lan, Xingyu and Wu, Yanqiu and Shi, Yang and Chen, Qing and Cao, Nan},
  booktitle={Proceedings of the 2022 CHI Conference on Human Factors in Computing Systems},
  pages={1--14},
  year={2022}
}

@article{lugmayr2017serious,
  title={Serious storytelling--a first definition and review},
  author={Lugmayr, Artur and Sutinen, Erkki and Suhonen, Jarkko and Sedano, Carolina Islas and Hlavacs, Helmut and Montero, Calkin Suero},
  journal={Multimedia tools and applications},
  volume={76},
  pages={15707--15733},
  year={2017},
  publisher={Springer}
}

@article{lee2015more,
  title={More than telling a story: Transforming data into visually shared stories},
  author={Lee, Bongshin and Riche, Nathalie Henry and Isenberg, Petra and Carpendale, Sheelagh},
  journal={IEEE computer graphics and applications},
  volume={35},
  number={5},
  pages={84--90},
  year={2015},
  publisher={IEEE}
}

@article{willett2021perception,
  title={Perception! immersion! empowerment! superpowers as inspiration for visualization},
  author={Willett, Wesley and Aseniero, Bon Adriel and Carpendale, Sheelagh and Dragicevic, Pierre and Jansen, Yvonne and Oehlberg, Lora and Isenberg, Petra},
  journal={IEEE transactions on visualization and computer graphics},
  volume={28},
  number={1},
  pages={22--32},
  year={2021},
  publisher={IEEE}
}

@article{lan2021smile,
  title={Smile or scowl? looking at infographic design through the affective lens},
  author={Lan, Xingyu and Shi, Yang and Zhang, Yueyao and Cao, Nan},
  journal={IEEE Transactions on Visualization and Computer Graphics},
  volume={27},
  number={6},
  pages={2796--2807},
  year={2021},
  publisher={IEEE}
}

@incollection{bach2018narrative,
  title={Narrative design patterns for data-driven storytelling},
  author={Bach, Benjamin and Stefaner, Moritz and Boy, Jeremy and Drucker, Steven and Bartram, Lyn and Wood, Jo and Ciuccarelli, Paolo and Engelhardt, Yuri and Koeppen, Ulrike and Tversky, Barbara},
  booktitle={Data-driven storytelling},
  pages={107--133},
  year={2018},
  publisher={AK Peters/CRC Press}
}

@article{denzin1983note,
  title={A note on emotionality, self, and interaction},
  author={Denzin, Norman K},
  journal={American journal of sociology},
  volume={89},
  number={2},
  pages={402--409},
  year={1983},
  publisher={University of Chicago Press}
}

@misc{COMICCop96:online,
author = {Jin, Connie Hanzhang},
title = {COMIC: Coping with pandemic numbness : Goats and Soda : NPR},
howpublished = {\url{https://www.npr.org/sections/goatsandsoda/2021/04/25/987208356/comic-how-i-cope-with-pandemic-numbness}},
year = {2021}
}

@misc{TheProje32:online,
author = {Grippe, John},
title = {The Project Behind a Front Page Full of Names - The New York Times},
howpublished = {\url{https://www.nytimes.com/2020/05/23/reader-center/coronavirus-new-york-times-front-page.html}},
year = {2020}
}

@misc{500000do29:online,
author = {Gagnon, Francis},
title = {500,000 dots is too many - Voilà:},
howpublished = {\url{https://chezvoila.com/blog/500k/}},
year = {2021}
}

@misc{Iraqsblo67:online,
author = {Scarr, Simon},
title = {Iraq's bloody toll — Simon Scarr},
howpublished = {\url{http://www.simonscarr.com/iraqs-bloody-toll/}},
year = {2011}
}

@article{purchase2002metrics,
  title={Metrics for graph drawing aesthetics},
  author={Purchase, Helen C},
  journal={Journal of Visual Languages \& Computing},
  volume={13},
  number={5},
  pages={501--516},
  year={2002},
  publisher={Elsevier}
}

@inproceedings{fogg2001makes,
  title={What makes web sites credible? A report on a large quantitative study},
  author={Fogg, Brian J and Marshall, Jonathan and Laraki, Othman and Osipovich, Alex and Varma, Chris and Fang, Nicholas and Paul, Jyoti and Rangnekar, Akshay and Shon, John and Swani, Preeti and others},
  booktitle={Proceedings of the SIGCHI conference on Human factors in computing systems},
  pages={61--68},
  year={2001}
}

@inproceedings{hartmann2006assessing,
  title={Assessing the attractiveness of interactive systems},
  author={Hartmann, Jan},
  booktitle={CHI'06 extended abstracts on Human factors in computing systems},
  pages={1755--1758},
  year={2006}
}

@inproceedings{kurosu1995apparent,
  title={Apparent usability vs. inherent usability: experimental analysis on the determinants of the apparent usability},
  author={Kurosu, Masaaki and Kashimura, Kaori},
  booktitle={Conference companion on Human factors in computing systems},
  pages={292--293},
  year={1995}
}

@article{scha1993computationele,
  title={Computationele esthetica},
  author={Scha, Remko and Bod, Rens},
  journal={Informatie en Informatiebeleid},
  volume={11},
  number={1},
  pages={54--63},
  year={1993}
}

@inproceedings{tractinsky1997aesthetics,
  title={Aesthetics and apparent usability: empirically assessing cultural and methodological issues},
  author={Tractinsky, Noam},
  booktitle={Proceedings of the ACM SIGCHI Conference on Human factors in computing systems},
  pages={115--122},
  year={1997}
}

@article{ngo2001another,
  title={Another look at a model for evaluating interface aesthetics},
  author={Ngo, David Chek Ling and Byrne, John G},
  journal={International Journal of Applied Mathematics and Computer Science},
  volume={11},
  number={2},
  pages={515--535},
  year={2001},
  publisher={Uniwersytet Zielonog{\'o}rski. Oficyna Wydawnicza}
}

@article{stasko2000evaluation,
  title={An evaluation of space-filling information visualizations for depicting hierarchical structures},
  author={Stasko, John and Catrambone, Richard and Guzdial, Mark and McDonald, Kevin},
  journal={International journal of human-computer studies},
  volume={53},
  number={5},
  pages={663--694},
  year={2000},
  publisher={Elsevier}
}

@article{chen2005top,
  title={Top 10 unsolved information visualization problems},
  author={Chen, Chaomei},
  journal={IEEE computer graphics and applications},
  volume={25},
  number={4},
  pages={12--16},
  year={2005},
  publisher={IEEE}
}

@inproceedings{sutcliffe2001heuristic,
  title={Heuristic evaluation of website attractiveness and usability},
  author={Sutcliffe, Alistair},
  booktitle={International workshop on design, specification, and verification of interactive systems},
  pages={183--198},
  year={2001},
  organization={Springer}
}

@article{overbeeke2004beauty,
  title={Beauty in use},
  author={Overbeeke, Kees and Wensveen, Stephan},
  journal={Human--Computer Interaction},
  volume={19},
  number={4},
  pages={367--369},
  year={2004},
  publisher={Taylor \& Francis}
}

@inproceedings{petersen2004aesthetic,
  title={Aesthetic interaction: a pragmatist's aesthetics of interactive systems},
  author={Petersen, Marianne Graves and Iversen, Ole Sejer and Krogh, Peter Gall and Ludvigsen, Martin},
  booktitle={Proceedings of the 5th conference on Designing interactive systems: processes, practices, methods, and techniques},
  pages={269--276},
  year={2004}
}

@inproceedings{cawthon2007effect,
  title={The effect of aesthetic on the usability of data visualization},
  author={Cawthon, Nick and Moere, Andrew Vande},
  booktitle={2007 11th International Conference Information Visualization (IV'07)},
  pages={637--648},
  year={2007},
  organization={IEEE}
}

@article{sayago2016conceptualization,
  title={On the conceptualization, design, and evaluation of appealing, meaningful, and playable digital games for older people},
  author={Sayago, Sergio and Rosales, Andrea and Righi, Valeria and Ferreira, Susan M and Coleman, Graeme W and Blat, Josep},
  journal={Games and Culture},
  volume={11},
  number={1-2},
  pages={53--80},
  year={2016},
  publisher={SAGE Publications Sage CA: Los Angeles, CA}
}

@misc{giblin2019dismantling,
  title={Dismantling the master’s house: thoughts on representing empire and decolonising museums and public spaces in practice an introduction},
  author={Giblin, John and Ramos, Imma and Grout, Nikki},
  journal={Third Text},
  volume={33},
  number={4-5},
  pages={471--486},
  year={2019},
  publisher={Taylor \& Francis}
}

@article{luebke1994locating,
  title={Locating the victim: An overview of census-taking, tabulation technology, and persecution in Nazi Germany},
  author={Luebke, David Martin and Milton, Sybil},
  journal={IEEE Annals of the History of Computing},
  volume={16},
  number={03},
  pages={25--39},
  year={1994},
  publisher={IEEE Computer Society}
}

@article{baines1983literacy,
  title={Literacy and ancient Egyptian society},
  author={Baines, John},
  journal={Man},
  pages={572--599},
  year={1983},
  publisher={JSTOR}
}

@article{tham2020systematic,
  title={A systematic investigation of conceptual color associations.},
  author={Tham, Diana Su Yun and Sowden, Paul T and Grandison, Alexandra and Franklin, Anna and Lee, Anna Kai Win and Ng, Michelle and Park, Juhyun and Pang, Weiguo and Zhao, Jingwen},
  journal={Journal of Experimental Psychology: General},
  volume={149},
  number={7},
  pages={1311},
  year={2020},
  publisher={American Psychological Association}
}

@article{bazley2021visual,
  title={Visual Finance:: The Pervasive Effects of Red on Investor Behavior},
  author={Bazley, William J and Cronqvist, Henrik and Mormann, Milica},
  journal={Management Science},
  volume={67},
  number={9},
  pages={5616--5641},
  year={2021},
  publisher={INFORMS}
}

\end{document}